\documentstyle[twocolumn,pra,aps,epsfig]{revtex}
\begin{document}
\flushbottom
\draft
\title{Quantum optics of a Bose-Einstein condensate coupled to
a quantized light field}
\author{M. G. Moore, O. Zobay, and P. Meystre}
\address{Optical Sciences Center and Department of Physics\\
University of Arizona, Tucson, Arizona 85721\\
(May 19, 1998)
\\ \medskip}\author{\small\parbox{14.2cm}{\small \hspace*{3mm}
We consider the interaction between a Bose-Einstein condensate and
a single-mode quantized light field in the presence of a strong
far off-resonant pump laser. The dynamics is characterized by an
exponential instability, hence the system acts as an atom-photon parametric
amplifier.  Triggered by a small injected probe field, or simply by quantum
noise, entangled atom-photon pairs are created which exhibit
non-classical correlations similar to those seen between photons in
the optical parametric amplifier. In addition, the quantum statistics
of the matter and light fields depend strongly on the initial state
which triggers the amplifier. Thus by preparing different initial states
of the light field, one can generate matter waves in a variety of 
quantum states, demonstrating optical control over the quantum statistics
of matter waves.
\\[3pt]PACS numbers: 03.75.-b,42.50.-p,42.50.Ct,42.50.Dv  }}
\maketitle
\narrowtext

\section{Introduction}

In many ways, recently developed Bose Einstein condensates (BEC) of trapped 
alkali atomic vapors \cite{AndEnsMat95,DavMewAnd95}
are the atomic analog of the optical laser. In fact, with
the addition of an output coupler, they are frequently referred to as
`atom lasers' \cite{MewAndKur97}. Despite many interesting and important 
differences, the chief similarity behind the analogy is that both optical 
lasers and atomic BEC's involve large numbers of identical bosons occupying
a single quantum state. As a result, the physics of lasers and BEC involves
stimulated processes, which due to Bose enhancement
often completely dominate the spontaneous processes which
play central roles in the non-degenerate regime. 

Just as the discovery of the laser led to the development
of nonlinear optics, so too has the advent of BEC led to remarkable
experimental successes in the once theoretical field of nonlinear atom optics
\cite{LenMeyWri93,LenMeyWri94,ZhaWalSan94,ZhaWal94,CasMol95}. Nonlinear optics typically
involves the study of multi-wave mixing,
epitomized by phenomena such as parametric down conversion and phase 
conjugation. Due to the presence of collisions, the evolution of the atomic
field is also nonlinear, and multi-wave mixing has been predicted
\cite{GolPlaMey95,GolPlaMey96,TriBanJul98,LawPuBig98,GolMey99a,GolMey99b}
and observed in multi-component condensates \cite{SteInoSta98}, as well as in
scalar condensates \cite{Phi4wm}.

At the root of most optical phenomena is the dynamical interaction between
optical and atomic fields. Under certain circumstances, one can formally
eliminate the dynamics of the atomic field, resulting in effective
interactions between light waves. Under a different set of conditions, one can 
eliminate the electromagnetic field dynamics, resulting in effective atom-atom
interactions. These are the regimes of nonlinear optics and nonlinear atom
optics, respectively. These regimes, therefore, represent limiting
cases, where either the atomic or optical field is 
not dynamically independent, and instead follows the other field in some
adiabatic manner which allows for its effective elimination. 

Outside of these two regimes the 
atomic and optical fields are dynamically independent, 
and neither field is readily eliminated. In this paper we investigate the
dynamics of coupled quantum degenerate atomic and optical fields
in this intermediate regime. In particular, 
we investigate a system which is analogous to the non-degenerate optical 
parametric amplifier (OPA) \cite{She84,WalMil94}. However, whereas the OPA involves the 
creation of correlated photon pairs, this system involves the generation of
correlated atom-photon pairs. The purpose of this paper is to develop
a detailed theory for the interaction of quantized atomic and optical fields,
with emphasis on the manipulation and control of their quantum statistics and 
the generation of quantum correlations and entanglement between matter and 
light waves.

The specific system we consider consists of a Bose-Einstein condensate driven by a strong
far off-resonant pump laser which interacts with a single mode of an optical
ring cavity counterpropagating with respect to the pump. The strong pump 
laser is treated in the usual manner as a 
classical, undepleted light field and furthermore it is assumed to be
detuned far enough away from resonance that spontaneous emission
may be safely neglected. The cavity field, henceforth referred to as
the `probe', is assumed to be weak relative to the pump, and is treated fully 
quantum mechanically as a dynamical variable. It is the dynamical interplay
between this probe field and the atomic field which is the subject of interest.
The pump serves as a sort of catalyst, inducing a strong atomic dipole moment,
thus significantly enhancing the atom-probe interaction.

Assuming that the probe field begins in or near the vacuum state, and the
atomic field consists initially of a trapped BEC, the initial
dynamics is dominated by a single process: the absorption of a pump photon by a condensate
atom followed by the emission of a probe photon. We remark that in the
far off-resonant configuration, this is a two-photon virtual transition
in which the excited atomic state population remains negligible.
Due to atomic recoil, the absorption/emission process
transfers the atom from the condensate ground state to a new state that is
shifted in momentum space by the two-photon recoil. This new state
constitutes a second condensate component, which can be considered as a
momentum side mode to the original condensate. \footnote{This new state
may or may not be in the same internal ground state as the original condensate,
depending on the polarizations of the pump and probe photons. If the magnetic
sublevels are different, then the transition would be termed a Raman transition,
if the sublevels are the same, than it may be thought of as Rayleigh scattering
or two-photon Bragg 
scattering \cite{KozDenHag99,SteInoChi99}. As the states are already distinguished by
their center-of-mass momentum states, to further distinguish them by an
additional quantum number would add nothing. Our model deals
specifically with the Bragg scheme, however, with only minimal modifications it
could be applied to the Raman scheme as well.} As this side mode is populated, 
it begins to interfere with
the original condensate, resulting in fringes \cite{AndTowMie97}. These fringes 
are seen by the 
pump and probe fields as a spatial density grating, which then enhances the
photon scattering process.

This interplay between interference fringes and scattering can act as a 
positive feedback mechanism, in which case the system is
unstable, and is characterized by exponential growth. 
Any small signal, including quantum noise, will be sufficient to
trigger the instability, resulting in the generation of exponentially growing
side mode and probe fields. Of course this exponential growth is eventually
reversed by high intensity effects, so that the long-time dynamics
is characterized by large-amplitude nonlinear oscillations. 

At the present time, we focus on the small-signal regime, characterized 
by exponentially growing fields. In this regime, we demonstrate that
the quantum state of the probe and side mode fields depends strongly
on the initial conditions, so that, e.g., by injecting a small coherent
light field into the probe, one can create an entirely different quantum
state than that generated from the amplification of quantum vacuum
fluctuations. The differences are manifested in both the quantum statistics
of the individual field modes, as well as in non-classical correlations
and entanglement between them.

This rest of this paper is organized as follows. Section II gives the background
and relates the current theory to previous works in a variety of fields.
Section III outlines the basic model for a quantized many-body atomic field
interacting with a strong classical pump laser and a quantized optical cavity
mode. In Sec. IV, coupled-mode equations are developed for the condensate and
its momentum side modes. These equations are then linearized in Sec. V, 
resulting in a three-mode model which is exactly
solvable. Section VI then discusses the exponential instability, with
emphasis on the effects of collisions. In Sec. VII the quantum statistics of 
the atomic and electric fields are investigated, and the extent to which they 
can be manipulated is determined. In Sec. VIII atom-photon entanglement is
discussed, including an examination of two-mode squeezing between atomic and
optical fields Lastly, section IX is a discussion and
conclusion, which includes estimates of the important physical parameters, as
well as potential experimental obstacles. 
 
\section{background}

The system we describe is in fact an extension into the ultracold regime
of the theoretical work of Bonifacio and coworkers on the Collective Atomic 
Recoil Laser (CARL) \cite{BonSal94,BonSalNar94}.
The original CARL theory  treated the atomic center-of-mass motion classically,
an approximation certainly valid for hot atoms, but not sufficient to describe
ultracold samples such as BEC's. Within this framework, the feedback mechanism 
which gives rise to the exponential instability in the CARL was outlined 
using a slightly different, but complementary physical picture, where the
classical atomic center-of-mass motion in 
the optical potential of the counterpropagating pump and probe fields is 
responsible for the grating formation. The theory was extended to the limit of 
zero temperature by assuming that all of the (classical) atoms begin from rest, 
leading to the discovery of the so-called `CARL cubic equation', which gives
the exponential growth rate of the instability in terms of the relevant system
parameters. Out of a desire to better understand the quantum statistics of the
probe field, an attempt at a quantum $T=0$ theory
was made \cite{Bon98}. However, this attempt explicitly assumed
that the wavefunctions of the individual atoms could be localized
in both momentum and position space to an extent which violates the 
Heisenberg uncertainty principle. Thus rather than being a true quantum theory, 
it still treated the atoms as following `classical' trajectories, but now with 
small `quantum' fluctuations included. 

Both the original classical CARL model, as well as the later
`quantum' model, fall within the ray-optics approximation 
for the atomic field. Clearly then, one would expect such models to break down
as soon as the atomic deBroglie wavelength becomes comparable to the
period of the optical potential formed by the pump and probe fields.
As the wavelength of the optical potential is twice the optical wavelength,
this breakdown should occur near the atomic recoil 
temperature, which for typical alkali atoms is on the order of microKelvins.
As subrecoil temperature atomic vapors are achieved routinely through a variety
of cooling techniques, a theory which properly treats the quantum motion 
of the atoms is required if one desires to investigate the behavior of the
CARL in this regime.

With the ultimate intent of extending the CARL theory into the BEC regime, so
that the unique coherence properties of condensates might be further
understood and exploited by the interaction with dynamical light fields, a
quantum model of the atomic motion was formulated \cite{MooMey98},
where it was confirmed that the ray-optics versions did indeed break
down for temperatures of the order of the recoil temperature or below. 
In fact, at $T=0$,
a second threshold for the existence of the exponential instability 
was discovered, occurring when the bunching process is overcome by matter-wave 
diffraction. For $T>T_R$, however, it was shown that the previous theories
make indistinguishable predictions from the quantum theory. We remark that
while in this work the atomic center of mass motion was treated quantum 
mechanically, the light fields were still treated
classically, hence predictions concerning the quantum statistics of either
the atomic or optical fields could not be made. 

A full quantum model of both the atomic and optical fields  
was recently outlined in \cite{MooMey99a}, where the subjects of manipulating
quantum statistics and atom-photon entanglement were first addressed.
The present paper is a detailed elaboration and extension of that 
work, including significant new physics. For example, utilizing
the familiar s-wave scattering approach of BEC theory, the effects of
atom-atom collisions are incorporated into the CARL theory for
the first time. Also, in an extension of the OPA analogy, the existence of 
two-mode squeezing is shown to occur between a condensate side mode and the 
probe optical field. The current approach also differs from earlier work 
in that the familiar
spontaneous symmetry breaking technique is no longer applied to the condensate.
Instead it is assumed that a condensate well below the critical temperature is 
better described by a number state than a coherent state, as recent work 
appears to demonstrate \cite{WilWei97,GroHol97}.

The fully quantum model is similar in many ways to a system studied
by Zeng and coworkers \cite{ZenLinZha95}, in which the principle of 
manipulating the quantum statistics of a condensate by its interaction with a 
quantized light field was first proposed. This paper, however, does not  
recognize the existence of unstable (exponential) solutions 
nor the fact that the system can be triggered from quantum noise. 
We note that the unstable (exponential) solutions, and the possibility to 
initiate them from quantum vacuum fluctuations are both crucial components of 
this present work.
Lastly, we mention the connection to recent work on matter-wave amplification by
Law and Bigelow \cite{LawBig98}, which also explores the interaction
between condensates and quantized light fields. In that work, however, the
light field is assumed to be heavily damped, thus allowing for its dynamical
elimination. As a result, only the properties of the atomic field are
studied in detail.

CARL theory, including the present version, is also closely related to the
theory of Recoil Induced Resonances (RIR) \cite{GuoBerDub92}, 
in which the effects of atomic recoil on the pump-probe spectroscopy of an 
atomic vapor is investigated. This theory  treats the atomic center-of-mass 
motion 
quantum mechanically. The probe field, however, is not typically treated as a 
dynamical variable. Hence, it does not include the effects of probe feedback,
which are necessary for exponential behavior. A detailed comparison of the RIR and CARL 
theories is given in \cite{Ber99}.      

\section{The basic model}

In this section we derive a fully quantized model of a  gas of 
bosonic two-level atoms which interact with a strong, classical, undepleted 
pump laser and a weak, quantized optical ring cavity mode, both of which are 
assumed to be tuned far away from atomic resonances. As a result, single-photon 
transitions between atomic internal ground and excited states
are highly non-resonant and the excited state population remains 
negligible. In this case, one can safely neglect the effects of spontaneous 
emission as well as the two-body dipole-dipole interaction. 

We must still, however, allow for
two-photon virtual transitions in the which the atomic internal state remains
unchanged, but due to recoil may result in a change in the atom's center-of-mass
motion. For example, an atom which absorbs a pump photon and emits a probe
photon experiences a recoil kick equal to the difference of the momenta of the
two photons (which for nearly counterpropagating pump and probe beams
is of the order of two optical momenta). These transitions, therefore, couple 
different states of the atomic center-of-mass motion. Due to the 
quadratic dispersion relation of the atoms, these transitions will in 
general be non-resonant. For very cold atoms, the resultant detunings are 
typically on the order of the recoil frequency, i.e. much smaller than the 
natural linewidth of the atomic transition, $\gamma_a$, whereas the one-photon 
transitions which we are neglecting have a detuning many orders of magnitude 
larger than $\gamma_a$. 

Our theory begins with the second-quantized Hamiltonian
\begin{eqnarray}
\hat{\cal H}&=&\hat{\cal H}_{atom}+\hat{\cal H}_{probe}
+\hat{\cal H}_{atom-probe}\nonumber\\
&+&\hat{\cal H}_{atom-pump}+\hat{\cal H}_{atom-atom},
\label{H}
\end{eqnarray}
where $\hat{\cal H}_{atom}$ and $\hat{\cal H}_{probe}$ give 
the free evolution of the atomic field and the probe mode respectively,  
$\hat{\cal H}_{atom-probe}$ and $\hat{\cal H}_{atom-pump}$ describe the 
dipole coupling between the atomic field and the probe mode and pump laser,
respectively, and $\hat{\cal H}_{atom-atom}$ contains the two-body
$s$-wave scattering collisions between ground state atoms. 

The free atomic Hamiltonian is given by
\begin{eqnarray}
\hat{\cal H}_{atom}&=&\int d^3{\bf r}\left[\hat{\mit\Psi}^\dag_g({\bf r})
\left(-\frac{\hbar^2}{2m}\nabla^2+V_g({\bf r})\right)
\hat{\mit\Psi}_g({\bf r})\right.\nonumber\\
&+&\left.\hat{\mit\Psi}^\dag_e({\bf r})
\left(-\frac{\hbar^2}{2m}\nabla^2+\hbar\omega_a+V_e({\bf r})\right)
\hat{\mit\Psi}_e({\bf r})\right],
\label{Hatom}
\end{eqnarray}
where $m$ is the atomic mass, $\omega_a$ is the atomic resonance frequency,
$\hat{\mit\Psi}_e({\bf r})$ and $\hat{\mit\Psi}_g({\bf r})$ are the
atomic field operators for excited and ground state atoms respectively,
and $V_g({\bf r})$ and $V_e({\bf r})$ are their respective trap potentials.  
The atomic field operators obey the usual bosonic equal time commutation
relations $[\hat{\mit\Psi}_j({\bf r}),\hat{\mit\Psi}^\dag_{j^\prime}({\bf r}^\prime)]
 = \delta_{j,j^\prime}\delta^3({\bf r}-{\bf r}^\prime)$, and
$[\hat{\mit\Psi}_j({\bf r}),\hat{\mit\Psi}_{j^\prime}({\bf r}^\prime)] =
[\hat{\mit\Psi}^\dag_j({\bf r}),\hat{\mit\Psi}^\dag_{j^\prime}({\bf r}^\prime)] = 0$,
where $j,j^\prime=\{e,g\}$.

The free evolution of the probe mode is governed by the Hamiltonian
\begin{equation}
\hat{\cal H}_{probe}=\hbar ck\hat{A}^\dag\hat{A},
\label{Hprobe}
\end{equation}
where $c$ is the speed of light, $k$ is the magnitude of the probe wave number
${\bf k}$, and $\hat{A}$ and $\hat{A}^\dag$ are the probe photon annihilation 
and creation operators,
satisfying the boson commutation relation $[\hat{A},\hat{A}^\dag]=1$. 
The probe wavenumber ${\bf k}$ must satisfy the periodic boundary condition 
of the ring cavity, $k=2\pi\ell/L$, where the integer $\ell$ is 
the longitudinal mode index, and $L$ is the length of the cavity.

The atomic and probe fields interact in the dipole approximation 
via the Hamiltonian
\begin{eqnarray}
\hat{\cal H}_{atom-probe}&=&
-i\hbar g \hat{A} \int d^3{\bf r}\hat{\mit\Psi}^\dag_e({\bf r})e^{i{\bf k}
\cdot{\bf r}}\hat{\mit\Psi}_g({\bf r})\nonumber\\
&+&H.c.,
\label{Hatom-probe}
\end{eqnarray}
where $g=d[ck/(2\hbar\epsilon_0 LS)]^{1/2}$ is the atom-probe coupling constant.
Here $d$ is the magnitude of the atomic dipole moment, and $S$ is the 
cross-sectional area of the 
probe mode in the vicinity of the atomic sample (where it is assumed to be
approximately constant across the length of the atomic sample).

In addition, the atoms are driven by a strong pump laser, which is treated
classically and assumed to remain undepleted.  The atom-pump interaction
Hamiltonian is given in the dipole approximation by
\begin{eqnarray}
\hat{\cal H}_{atom-pump}&=&\frac{\hbar\Omega_0}{2}e^{-i\omega_0t}
\int d^3{\bf r}\hat{\mit\Psi}^\dag_e({\bf r})e^{i{\bf k}_0\cdot{\bf r}}
\hat{\mit\Psi}_g({\bf r})\nonumber\\
&+& H.c.,
\label{Hatom-pump}
\end{eqnarray}
where $\Omega_0$ is the Rabi frequency of the pump laser, related to the
pump intensity $I_0$ by $|\Omega_0|^2=2d^2I_0/\hbar^2\epsilon_0c$,
$\omega_0$ is the pump frequency, and $k_0\approx \omega_0/c$ is the pump 
wavenumber. The approximation indicates that we are neglecting the
index of refraction inside the atomic gas, as we assume a very large
detuning $\Delta=\omega_0-\omega_a$ between the pump frequency and the
atomic resonance frequency.

Finally, the collision Hamiltonian is taken to be
\begin{equation}
\hat{\cal H}_{atom-atom}=\frac{2\pi\hbar^2\sigma}{m}
\int d^3{\bf r}\hat{\mit\Psi}^\dag_g({\bf r})\hat{\mit\Psi}^\dag_g({\bf r})
\hat{\mit\Psi}_g({\bf r})\hat{\mit\Psi}_g({\bf r}),
\label{Hatom-atom}
\end{equation}
where $\sigma$ is the atomic $s$-wave scattering length.
This corresponds to the usual $s$-wave scattering approximation, and 
leads in the Hartree approximation to the standard Gross-Pitaevskii equation 
for the ground state wavefunction 
(in the absence of the driving optical fields).

We limit ourselves to the case where the pump laser is detuned
far enough away from the atomic resonance that the excited state population
remains negligible, a condition which requires that $\Delta\gg\gamma_a$. In 
this regime the atomic polarization adiabatically follows the
ground state population, allowing the formal elimination of the
excited state atomic field operator. We proceed by introducing the
operators $\hat{\mit\Psi}^\prime_e({\bf
r})=\hat{\mit\Psi}_e({\bf r})e^{i\omega_0t}$ and
$\hat{A}^\prime=\hat{A}e^{i\omega_0t}$, which are slowly varying relative to
the optical driving frequency.  The new excited state
atomic field operator obeys then the Heisenberg equation of motion
\begin{eqnarray}
\frac{d}{dt}\hat{\mit\Psi}^\prime_e({\bf r})&=&i\Delta\hat{\mit\Psi}^\prime_e({\bf r})
-\left[i\frac{\Omega_0}{2}e^{i{\bf k}_0\cdot{\bf r}}
+g\hat{A}^\prime e^{i{\bf k}\cdot{\bf r}}\right]\hat{\mit\Psi}_g({\bf r}),
\label{exciteq}
\end{eqnarray} 
where we have dropped the kinetic energy and trap potential terms under the 
assumption that
the lifetime of the excited atom, which is of the order $1/\Delta$, is so small
that atomic center-of-mass motion may be safely neglected during this period.
For the same reason, we are justified in neglecting collisions between
excited atoms, or between excited and ground state atoms in the collision
Hamiltonian (\ref{Hatom-atom}).
  
We now adiabatically solve for  $\hat{\mit\Psi}^\prime_e({\bf r})$ by 
formally integrating Eq. (\ref{exciteq}) under the 
assumption that $\hat{\mit\Psi}_g({\bf r})$ varies on a time scale
which is much longer than $1/\Delta$. This yields
\begin{eqnarray}
\hat{\mit\Psi}^\prime_e({\bf r},t)&\approx&
\frac{1}{\Delta}\left[\frac{\Omega_0}{2}
e^{i{\bf k}_0\cdot{\bf r}}-ig\hat{A}^\prime(t) e^{i{\bf k}\cdot{\bf r}}\right]
\hat{\mit\Psi}_g({\bf r},t)\nonumber\\
&-&\frac{1}{\Delta}\left[\frac{\Omega_0}{2}
e^{i{\bf k}_0\cdot{\bf r}}-ig\hat{A}^\prime(0) e^{i{\bf k}\cdot{\bf r}}\right]
\hat{\mit\Psi}_g({\bf r},0)e^{i\Delta t}\nonumber\\
&+&\hat{\mit\Psi}^\prime_e({\bf r},0)
e^{i\Delta t}.
\label{adelim}
\end{eqnarray}
The third term on the r.h.s. of Eq. (\ref{adelim}) can be neglected for most
considerations if we assume that there are no excited atoms at $t=0$, so that
this term acting on the initial state gives zero. The second term may also
be neglected, as it is rapidly oscillating at frequency $\Delta$, and thus its 
effect on the ground state field operator is
negligible when compared to that of the first term, which is non-rotating.
\footnote{We note that in much of the literature the second and third terms
are simply ignored. We choose to keep them temporarily to demonstrate
that the commutation relation for $\hat{\mit\Psi}_e({\bf r})$
is preserved (to order $1/\Delta$) by the procedure of adiabatic 
elimination.}
 
Dropping the unimportant terms, and then substituting Eq. (\ref{adelim}) 
into the equation of motion for $\hat{\mit\Psi}_g({\bf r})$, we arrive at the 
effective Heisenberg equation of motion for the ground state field operator
\begin{eqnarray}
\frac{d}{dt}\hat{\mit\Psi}_g({\bf r})&=&i\left[\frac{\hbar}{2m}\nabla^2
-\frac{V_g({\bf r})}{\hbar}
-\frac{4\pi\hbar\sigma}{m}\hat{\mit\Psi}^\dag_g({\bf r})
\hat{\mit\Psi}_g({\bf r})\right.\nonumber\\
&-&\frac{g|\Omega_0|}{2|\Delta|}\left(\hat{a}e^{i{\bf K}\cdot{\bf r}}
+\hat{a}^\dag e^{-i{\bf K}\cdot{\bf r}}\right)\nonumber\\
&-&\left.\left(\frac{|\Omega_0|^2}{4\Delta}
+\frac{g^2}{\Delta}\hat{a}^\dagger\hat{a}\right)\right]
\hat{\mit\Psi}_g({\bf r}),
\label{dPsidt}
\end{eqnarray}
where ${\bf K}={\bf k}-{\bf k}_0$ is the recoil momentum kick the
atom acquires from the two-photon transition, and we have introduced
the new slowly-varying probe field operator
$\hat{a}=-i(\Omega^\ast_0\Delta/|\Omega_0||\Delta|)\hat{A}^\prime$, which still
obeys the boson commutation relation $[\hat{a},\hat{a}^\dag]=1$.
Here, the second to last term is simply the optical potential formed from the
counterpropagating pump and probe light fields, and the last term gives the
spatially independent light shift potential, which can be thought of as
cross-phase modulation between the atomic and optical fields.

To complete our model, in addition to Eq. (\ref{dPsidt}), we
also require the equation of motion for the slowly varying probe field
operator. By again substituting Eq. (\ref{adelim}), we find that
it obeys
\begin{equation}
\frac{d}{dt}\hat{a}=i\delta^\prime\hat{a}
-i\frac{g|\Omega_0|}{2|\Delta|}\int d^3{\bf r}
\hat{\mit\Psi}^\dag_g({\bf r})e^{-i{\bf K}\cdot{\bf r}}
\hat{\mit\Psi}_g({\bf r}),
\label{dadt}
\end{equation}
where $\delta^\prime=\omega_0-\omega$, is the detuning between the pump and
probe fields. The probe frequency is given by $\omega\approx ck$, again 
assuming that the index of refraction inside the condensate is negligible.
 
\section{Coupled-mode equations}

We assume that the atomic field is initially in a 
Bose-Einstein condensate with mean number of condensed atoms $N$. 
Furthermore, we assume that $N$ is very large. 
and that the condensate temperature is small compared to the critical
temperature. These assumptions
allow us to neglect the noncondensed fraction of the atomic 
field. Thus our model does not include any effects of 
condensate number fluctuations.  
 
We now introduce the atomic field operator
which annihilates an atom in the condensate ground state
\begin{equation}
\hat{c}_0=\int d^3{\bf r}\varphi^\ast_0({\bf r})
\hat{\mit\Psi}_g({\bf r}),
\label{defc0}
\end{equation}
where $\varphi_0({\bf r})=\langle{\bf r}|\varphi_0\rangle$ satisfies the
time-independent Gross-Pitaevskii equation
\begin{equation}
\left(\frac{\hbar}{2m}\nabla^2-\frac{V_g({\bf r})}{\hbar}
-\frac{4\pi\hbar\sigma}{m}N|\varphi_0({\bf r})|^2+\frac{\mu}{\hbar}\right)
\varphi_0({\bf r})=0,
\label{GPE}
\end{equation}
$\mu$ being the chemical potential.
By differentiating Eq. (\ref{defc0}) with respect to time, and
inserting Eqs. (\ref{dPsidt}) and (\ref{GPE}) we find that the equation of 
motion for $\hat{c}_0$ is
\begin{eqnarray}
& &\frac{d}{dt}\hat{c}_0=-i\left(\frac{\mu}{\hbar}+\frac{|\Omega_0|^2}{4\Delta}
+\frac{g^2}{\Delta}\hat{a}^\dag\hat{a}\right)\hat{c}_0\nonumber\\
& &+i\frac{4\pi\hbar\sigma}{m}\int d^3{\bf r}\varphi^\ast_0({\bf r})
\Big( N|\varphi_0({\bf r})|^2-\hat{\mit\Psi}^\dag_g({\bf r})
\hat{\mit\Psi}_g({\bf r})\Big) 
\hat{\mit\Psi}_g({\bf r})\nonumber\\
& &-i\frac{g|\Omega_0|}{2|\Delta|}
\int d^3{\bf r}\varphi^\ast_0({\bf r})
\left(\hat{a}e^{i{\bf K}\cdot{\bf r}}
+\hat{a}^\dag e^{-i{\bf K}\cdot{\bf r}}\right)
\hat{\mit\Psi}_g({\bf r}).
\label{dc0dt}
\end{eqnarray}
From this equation we see that the effect of the optical fields is to 
couple the condensate mode to two side modes, whose wavefunctions are given by
\begin{equation}
\langle{\bf r}|\varphi_\pm\rangle=\varphi_0({\bf r})e^{\pm i{\bf K}\cdot{\bf r}}.
\label{defvarphipm1}
\end{equation}
In principle, the collision
term in Eq. (\ref{dc0dt}) also couples the condensate mode to various 
neighboring modes. However, to be consistent with the assumption of a
pure condensate ($T=0$) we assume that collisions
alone do not populate any new atomic states. 

Defining the field operators for the (first-order) condensate side modes as
\begin{equation}
\hat{c}_\pm=\int d^3{\bf r}\langle\varphi_\pm|{\bf r}\rangle
\hat{\mit\Psi}_g({\bf r}),
\label{defcpm1}
\end{equation}
allows us to reexpress the equation of motion for the
condensate mode field operator as
\begin{eqnarray}
& &\frac{d}{dt}\hat{c}_0=-i\left(\frac{\mu}{\hbar}+\frac{|\Omega_0|^2}{4\Delta}
+\frac{g^2}{\Delta}\hat{a}^\dag\hat{a}\right)\hat{c}_0\nonumber\\
& &+i\frac{4\pi\hbar\sigma}{m}\int d^3{\bf r}\varphi^\ast_0({\bf r})
\Big[ N|\varphi_0({\bf r})|^2-\hat{\mit\Psi}^\dag_g({\bf r})
\hat{\mit\Psi}_g({\bf r})\Big] 
\hat{\mit\Psi}_g({\bf r})\nonumber\\
& &-i\frac{g|\Omega_0|}{2|\Delta|}
(\hat{a}\hat{c}_-+\hat{a}^\dag\hat{c}_+),
\label{dc0dt2}
\end{eqnarray}
where the operators $\hat{c}_j$
obey the bosonic commutation relations
\begin{equation}
[\hat{c}_j,\hat{c}^\dag_{j^\prime}]=\langle\varphi_j|\varphi_{j^\prime}\rangle;
\quad j,j^\prime=\{-,0,+\},
\label{commutator1}
\end{equation}
all other commutators being equal to zero.

We note that the three states, $|\varphi_0\rangle$, and $|\varphi_\pm\rangle$ are not 
mutually orthogonal, as their overlap integrals are given by
\begin{eqnarray} 
\langle\varphi_\mp|\varphi_\pm\rangle&=&\int d^3{\bf r}
|\varphi_0({\bf r})|^2e^{\pm i2{\bf K}\cdot{\bf r}}\nonumber\\
\langle\varphi_0|\varphi_\pm\rangle&=&\int d^3{\bf r}
|\varphi_0({\bf r})|^2e^{\pm i{\bf K}\cdot{\bf r}}.
\label{overlap}
\end{eqnarray}
For most condensate sizes and trap configurations, however, these 
integrals are many orders of magnitude smaller than unity.
As a result, for `typical' condensates, the orthogonality approximation
\begin{equation}
\langle\varphi_j|\varphi_{j^\prime}\rangle=\delta_{jj^\prime}
\label{commutator2}
\end{equation}
yields accurate results.
The range of validity of this approximation is discussed in Appendix A, 
where we briefly examine how the theory should be modified to properly 
take this non-orthogonality into account. In the following, however, we assume the 
validity of Eq. (\ref{commutator2}), so that the states $|\varphi_0\rangle$,
and $|\varphi_\pm\rangle$ can be considered as well defined and distinct
modes of the atomic field.

We now derive the Heisenberg equations for the momentum side mode field
operators, found by differentiating Eq. (\ref{defcpm1}) with respect to time and
again inserting Eq. ({\ref{dPsidt}), yielding
\begin{eqnarray}
& &\frac{d}{dt}\hat{c}_{-,+}=-i\left(\frac{\mu}{\hbar}+\frac{\hbar K^2}{2m}
+\frac{|\Omega_0|^2}{4\Delta}+\frac{g^2}{\Delta}\hat{a}^\dag\hat{a}
\right)\hat{c}_{-,+}\nonumber\\
& &+i\frac{4\pi\hbar\sigma}{m}\int d^3{\bf r}
\langle\varphi_{-,+}|{\bf r}\rangle
\Big( N|\varphi_0({\bf r})|^2-\hat{\mit\Psi}^\dag_g({\bf r})
\hat{\mit\Psi}_g({\bf r})\Big) \hat{\mit\Psi}_g({\bf r})\nonumber\\
& &-i\frac{g|\Omega_0|}{2|\Delta|}\left(\hat{a}^\dag\hat{c}_{0,+2}
+\hat{a}\hat{c}_{-2,0}\right)+i\frac{\hbar Kk_c}{m}\hat{b}_{-,+},
\label{dcpmdt}
\end{eqnarray}
where we have introduced four new field operators $\hat{c}_{\pm 2}$ and 
$\hat{b}_\pm$.

The operators $\hat{c}_{\pm 2}$, which have the definitions 
\begin{equation}
\hat{c}_{\pm 2}=\int d^3{\bf r}\langle\varphi_{\pm 2}|{\bf
r}\rangle\hat{\mit\Psi}_g({\bf r}),
\label{defcm2}
\end{equation}
are the annihilation operators for the second-order side modes
\begin{equation}
\langle{\bf r}|\varphi_{\pm 2}\rangle=\varphi_0({\bf r})
e^{\pm i2{\bf K}\cdot{\bf r}},
\label{defvarphipm2}
\end{equation}
These modes will be optically coupled to third-order side modes,
and so on so that a full theory of the nonlinear response of the system
should include the entire manifold of side modes. In this paper, however, 
we focus on the
linear regime, where only the first-order side modes contribute significantly.

The operators $\hat{b}_\pm$ have the definitions 
\begin{equation}
\hat{b}_\pm=\int d^3{\bf r}\langle\phi_\pm|{\bf r}\rangle
\hat{\mit\Psi}_g({\bf r}),
\label{defbpm}
\end{equation}
where
\begin{equation}
\langle{\bf r}|\phi_\pm\rangle=(Kk_{rms})^{-1}e^{\pm i{\bf K}\cdot{\bf r}}
\big(\pm i{\bf K}\cdot{\bf \nabla}\varphi_0({\bf r})\big).
\label{defphipm}
\end{equation}
Here $k_c$ is the momentum width of the condensate state along ${\bf K}$,
and is roughly given by $k_c\sim 1/W_c$, where $W_c$ is the size of the
condensate along ${\bf K}$.
The factor $(Kk_c)^{-1}$ is simply a normalization coefficient.

To understand the physical meaning of the $\hat{b}_\pm$ term in Eq.
(\ref{dcpmdt}), consider what happens to a single atom after it is
transferred into the state $\psi({\bf r})=\varphi_0({\bf
r})\exp(-i{\bf K}\cdot{\bf r})$ at time $t=0$. Under free evolution 
the wavepacket of the atom, which initially has the shape of the condensate
ground state, will move with group velocity $\hbar{\bf K}/m$ and spread at
the velocity $\hbar k_c/m$. This evolution is described by the propagation
equation
\begin{equation}
\psi({\bf r},t)=\exp[it(\hbar/2m)\nabla^2]\psi({\bf r},0),
\label{prop}   
\end{equation}
which for short enough times becomes
\begin{eqnarray}
\psi({\bf r}, t)&\approx&\left(1-it\frac{\hbar K^2}{2m}\right)\langle{\bf
r}|\varphi_-\rangle
+t\frac{\hbar}{m}\big({\bf K}\cdot{\bf \nabla}
\varphi_0({\bf r})\big)e^{-i{\bf K}\cdot{\bf r}}\nonumber\\
&+&it\frac{\hbar}{2m}\big(\nabla^2\varphi_0({\bf r})\big)e^{-i{\bf K}
\cdot{\bf r}}.
\label{prop2}
\end{eqnarray}
The first term on the r.h.s. of Eq. (\ref{prop2}) gives a phase shift due to the kinetic energy of 
the atom, the second term contributes an infinitesimal translational shift,
and the third term gives an infinitesimal amount of spreading.
If we include the effects of the trap potential and collisions, this
last term vanishes as all spreading effects are balanced by the trap potential 
for the ground state $\varphi_0({\bf r})$.
From Eqs. (\ref{prop2}) and (\ref{defphipm}) we see that the state of the atom at time $t$
can then be viewed as a coherent superposition of the state $|\varphi_-\rangle$
and the state $|\phi_-\rangle$. Thus the coupling to $\hat{b}_-$ in Eq.
(\ref{dcpmdt}) corresponds physically to translational motion of the side mode
wavepacket at the recoil velocity $v_r=\hbar K/m$. Since the probability at time
$t$ that the atom is still in the ground state 
is simply the overlap between $\psi({\bf r},t)$ and $\psi({\bf r},0)$,
it is clear that for times $t\ll W_c/v_r$ this probability will be essentially 
unity, and the coupling to $\hat{b}_-$ can be ignored.  

\section{Linearized three-mode model}

From Eq. (\ref{dcpmdt}), we see that the first-order
side modes are optically coupled to both the condensate mode and to second-order
side modes. For times short enough that the condensate is not significantly
depleted, the coupling back into the condensate is subject to Bose
enhancement due to the presence of $\sim N$ identical bosons in this mode.
The coupling to the second-order side mode, in contrast, is not
enhanced. Hence for these time scales, the higher-order side modes are not
expected to play a significant role. 
In addition, we consider only times $t\ll W_c/v_r$, so that
the translational coupling can be neglected.  

These arguments suggest developing an approach where the three
atomic field operators $\hat{c}_0$, $\hat{c}_-$, and $\hat{c}_+$ play a 
predominant role. Therefore, we expand the atomic field operator as
\begin{equation}
\hat{\mit\Psi}_g({\bf r})=
\langle{\bf r}|\varphi_0\rangle\hat{c}_0
+\langle{\bf r}|\varphi_-\rangle\hat{c}_-
+\langle{\bf r}|\varphi_+\rangle\hat{c}_+
+ \hat{\psi}({\bf r}),
\label{modeexpand}
\end{equation}
where the field operator $\hat{\psi}({\bf r})$ acts only on the 
orthogonal complement to the subspace spanned by the state vectors
$|\varphi_0\rangle$, $|\varphi_-\rangle$, and $|\varphi_+\rangle$. As a result,
$\hat{\psi}({\bf r})$ commutes with the creation operators for the three 
central modes.

In the next step, we use Eq. (\ref{modeexpand}) to
expand the atomic polarization and collision terms in Eqs. (\ref{dadt}),
(\ref{dc0dt2}), and (\ref{dcpmdt}), with the eventual goal of
deriving a closed set of operator equations which fully describes the
system dynamics. At present, we are considering four dominant modes,
the condensate and first-order side modes, as well as 
the optical probe mode. In the linear regime, however, we will see that 
the condensate mode can be dynamically eliminated, resulting in an effective 
three-mode model. 

In expanding the polarization and collision terms by means of Eq.
(\ref{modeexpand}), there are two principal considerations in determining
which are the dominant terms.
The first is Bose enhancement, which, in the regime of negligible
condensate depletion, strongly selects transitions involving the 
condensate mode. In order to estimate this effect, we assign a weight of 
$\sqrt{N}$ for each occurrence of the operators $\hat{c}_0$ and $\hat{c}^\dag_0$
in a given term. The second consideration is momentum conservation, 
which comes from the spatial integration in the polarization and collision
terms. Integrals over slowly varying functions such as $|\varphi_0({\bf r})|^2$,
or $|\varphi_0({\bf r})|^4$ are `momentum selected' and dominate over
integrals of rapidly oscillating functions such as $|\varphi_0({\bf
r})|^2\exp(-i{\bf K}\cdot{\bf r})$. 

With this approach we find that the equation of motion for the probe field
operator (\ref{dadt}) becomes
\begin{equation}
\frac{d}{dt}\hat{a}=i\delta^\prime\hat{a}-i\frac{g|\Omega_0|}{2|\Delta|}
\left[\hat{c}^\dag_-\hat{c}_0+\hat{c}^\dag_0\hat{c}_+\right].
\label{dadt2}
\end{equation}
Thus we see that the probe annihilation operator is coupled to the bilinear 
atomic field operators $\hat{c}^\dag_-\hat{c}_0$
and $\hat{c}^\dag_0\hat{c}_+$. 
These operators correspond physically to interference fringes, i.e. a periodic 
modulation of the atomic
density, which appear because the atoms are in a
coherent superposition of the side mode and condensate states. Gain in the 
probe can thus be interpreted as Bragg scattering
of the pump due to the presence of interference fringes.  

By inserting (\ref{modeexpand}) into Eq. (\ref{dc0dt2}) we further find that
the equation of motion for $\hat{c}^\dag_0\hat{c}_0$ is given to leading
order in the collision and optical terms by
\begin{eqnarray}
\frac{d}{dt}\hat{c}^\dag_0\hat{c}_0&=&i\frac{8\pi\hbar\sigma F_0}{m}
\hat{c}^\dag_-\hat{c}_0\hat{c}^\dag_+\hat{c}_0
+i\frac{g|\Omega_0|}{2|\Delta|}\hat{a}^\dag(\hat{c}^\dag_-\hat{c}_0
-\hat{c}^\dag_0\hat{c}_+)\nonumber\\
&+&H.c.,
\label{dc0c0dt}
\end{eqnarray}
where 
\begin{equation}
F_0=\int d^3{\bf r}|\varphi_0({\bf r})|^4.
\label{defFj}
\end{equation}
Similarly we find that the operators $\hat{c}^\dag_-\hat{c}_0$ and
$\hat{c}^\dag_0\hat{c}_+$ obey the equations
\begin{eqnarray}
\frac{d}{dt}\hat{c}^\dag_-\hat{c}_0&=&
i\frac{\hbar K^2}{2m}\hat{c}^\dag_-\hat{c}_0
+i\frac{4\pi\hbar\sigma F_0}{m}
(\hat{c}^\dag_-\hat{c}_0+\hat{c}^\dag_0\hat{c}_+)
\hat{c}^\dag_0\hat{c}_0\nonumber\\
&+&i\frac{g|\Omega_0|^2|}{2|\Delta|}\hat{a}\hat{c}^\dag_0\hat{c}_0,
\label{dcm1c0dt}
\end{eqnarray}
and
\begin{eqnarray}
\frac{d}{dt}\hat{c}^\dag_0\hat{c}_+&=&
-i\frac{\hbar K^2}{2m}\hat{c}^\dag_0\hat{c}_+
-i\frac{4\pi\hbar\sigma F_0}{m}
(\hat{c}^\dag_-\hat{c}_0+\hat{c}^\dag_0\hat{c}_+)
\hat{c}^\dag_0\hat{c}_0\nonumber\\
&-&i\frac{g|\Omega_0|^2}{2|\Delta|}\hat{a}\hat{c}^\dag_0\hat{c}_0.
\label{dc0cp1dt}
\end{eqnarray}

We assume that all $N$ atoms are initially in the
condensate mode, so that
\begin{equation}
|\psi\rangle_{t=0}=\frac{1}{\sqrt{N!}}\left(\hat{c}^\dag_0\right)^N|0\rangle,
\label{initialstate}
\end{equation}
$|0\rangle$ being the vacuum state.
We proceed by linearizing the atomic field operators around their initial 
expectation values,
which can be determined from Eq. (\ref{initialstate}), together with the approximate
commutation relations given by Eq. (\ref{commutator2}). This yields 
\begin{equation}
\hat{c}^\dag_0\hat{c}_0= N(1+\hat{\delta}_0),
\label{defdeltaN}
\end{equation}
\begin{equation}
\hat{c}^\dag_-\hat{c}_0= N\hat{\delta}_-,
\label{defdeltam}
\end{equation}
and
\begin{equation}
\hat{c}^\dag_0\hat{c}_+= N\hat{\delta}_+,
\label{defdeltap}
\end{equation}
where $\hat{\delta}_0$, $\hat{\delta}_-$ and $\hat{\delta}_+$
are therefore infinitesimal operators.
In addition, we introduce a rescaled probe field operator
\begin{equation}
\hat{\delta}_a=\frac{\hat{a}}{\sqrt{N}},
\label{defdeltaa}
\end{equation}
which, provided that the mean number of 
photons in the probe mode is small compared to $N$, is also infinitesimal. 
This constraint is consistent with the assumption of negligible 
condensate depletion.

Inserting these definitions into the equation of motion (\ref{dc0c0dt}) for
$\hat{c}^\dag_0\hat{c}_0$, and keeping only terms linear in the infinitesimal 
operators, gives
\begin{equation}
\frac{d}{dt}\hat{\delta}_0=0,
\label{linddela0dt}
\end{equation}
which has the trivial solution $\hat{\delta}_0=0$.  
As a result, this operator
can be dropped from the linearized equations for $\hat{\delta}_-$ and
$\hat{\delta}_+$. This leads to a set of three coupled infinitesimal operators
whose linearized equations of motion can be expressed as
\begin{equation}
\frac{d}{d\tau}\hat{\vec{\delta}}=i{\bf M}\hat{\vec{\delta}},
\label{ddeltadt}
\end{equation}
where $\hat{\vec{\delta}}=(\hat{\delta}_a,
\hat{\delta}_-,\hat{\delta}_+)^T$, the matrix ${\bf M}$ is given by
\begin{equation}
{\bf M}=\left(\matrix{
\delta&-\chi&-\chi\cr
\chi&(1+\beta)&\beta\cr
-\chi&-\beta&-(1+\beta)\cr
}\right),
\label{defM}
\end{equation}
and we have introduced the dimensionless time $\tau=\omega_r t$, $\omega_r=
\hbar K^2/2m$ being the atomic recoil frequency, as well as the dimensionless
control parameters 
\begin{equation}
\chi=\frac{g|\Omega_0|\sqrt{N}}{2|\Delta|\omega_r},
\label{defchi}
\end{equation}\begin{equation}
\delta=\delta^\prime/\omega_r,
\label{defdelta}
\end{equation}
and
\begin{equation}
\beta=\frac{4\pi\hbar\sigma N F_0}{m\omega_r}.
\label{defbeta}
\end{equation}
Here $\chi$ is a dimensionless atom-probe coupling 
constant, $\delta$ is the pump-probe detuning in units of the atomic recoil
frequency, and $\beta$ gives the strength of collisions 
between the side modes. In Appendix B we give the effective Hamiltonian
from which Eq. (\ref{ddeltadt}) can be derived.

The solution to Eq. (\ref{ddeltadt}) is then given by
\begin{equation}
\hat{\vec{\delta}}(\tau)=e^{i{\bf M}\tau}\hat{\vec{\delta}}(0).
\label{deltaoft}
\end{equation}
From this we see that the time dependence of the infinitesimal operators
is determined by the eigenvalue spectrum of the matrix ${\bf M}$.
For certain values of the control
parameters $\chi$, $\delta$, and $\beta$, one of the eigenvalues
contains a negative imaginary part. When this occurs, the
infinitesimal operators undergo exponential growth.
This exponential instability is the focus of the next section,
where we discuss its properties in detail. 

\section{Exponential instability}

The eigenvalues of ${\bf M}$ are determined by the characteristic
equation
\begin{equation}\label{chareq}
\omega^3 - \delta\omega^2-(1+2\beta)\omega+(1+2\beta)\delta+2\chi^2=0,
\end{equation}
which has either three real solutions, or one real 
and a pair of complex conjugate solutions. In the first case, the system is 
stable and exhibits only small oscillations around its initial state. 
In the second case, the system is unstable and grows exponentially, even from
noise. 

From Eq.\ (\ref{chareq}) one finds that exponential instability
occurs when
\begin{equation}
\chi^2>[(3+6\beta+\delta^2)^{3/2}+\delta^3-9\delta(1+2\beta)]/27.
\label{expcond}
\end{equation}
In Fig. 1 we plot the region of instability as a function of
$\delta$ and $\chi^2$. The shaded region of Fig.1 corresponds to the
instability region in the absence of collisions $(\beta=0)$. As collisions
are added, the boundaries shift, illustrated by the dashed and dotted 
curves, which show the boundaries for the cases $\beta=.3$ and $\beta=1.0$, 
respectively.

From Fig. 1, we see that for positive $\delta$ the boundary
asymptotically reduces to 
\begin{equation}
\chi^2>2\delta^3/27-(1+2\beta)\delta/6
\label{+asympt}
\end{equation}
i.e., the threshold value of $\chi^2$ increases with the third power of the 
detuning and is only weakly influenced 
by the presence of collisions. On the other hand, 
for negative $\delta$ the asymptotic behavior is given by  
\begin{equation}
\chi^2>-(1+2\beta)\delta/2,
\label{-asymptote}
\end{equation}
which only grows linearly with $\delta$ and is strongly
affected by interatomic collisions, which have the effect of reducing
the unstable region. In earlier work \cite{MooMey98}, it was shown that 
this lower threshold occurs in the absence of collisions when atomic diffraction
overcomes the bunching process. For positive 
scattering lengths, 
\centerline{\psfig{figure=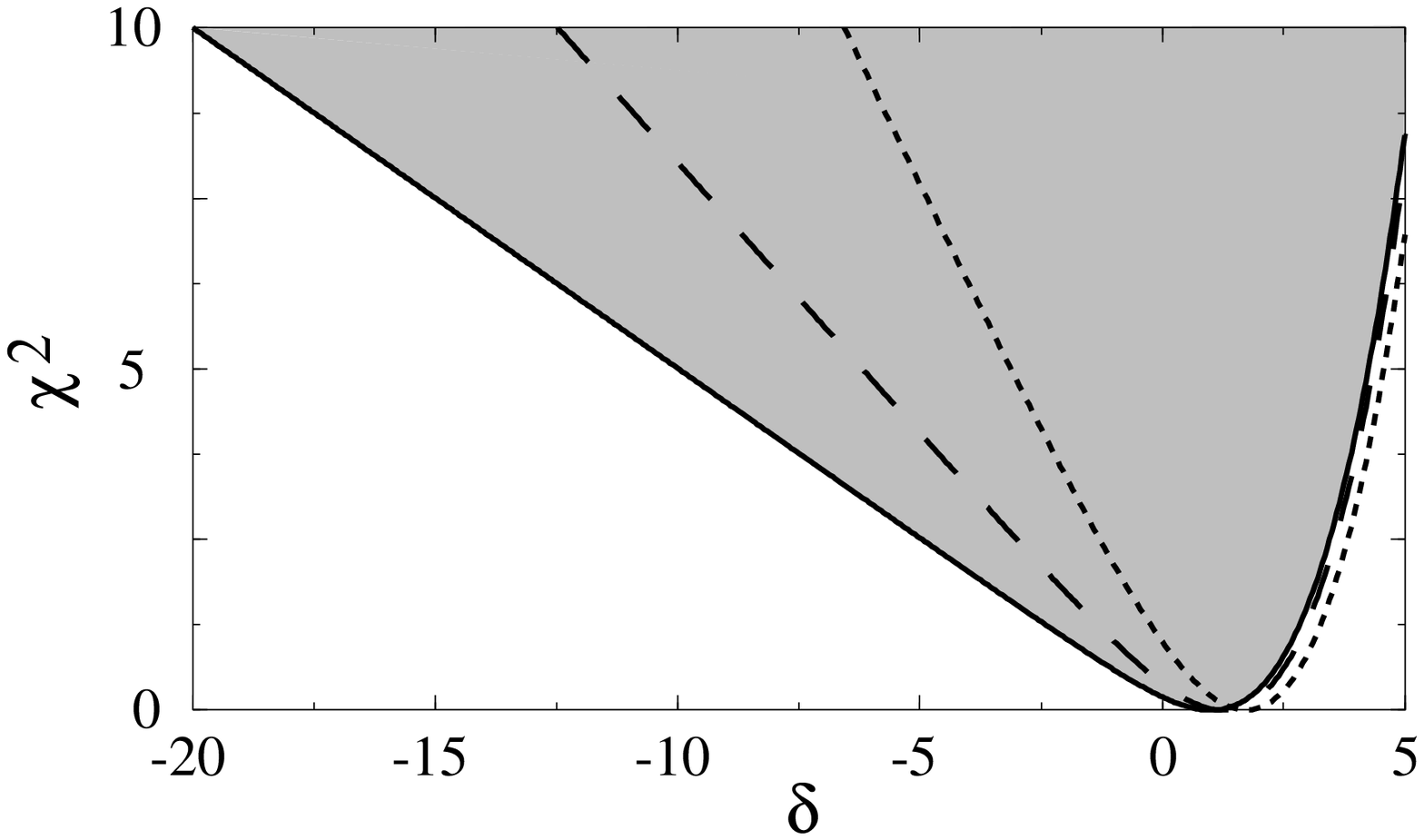,width=8.6cm,clip=1}}
\begin{figure}
\caption{Exponential instability region in the $\delta$-$\chi^2$ plane.
The shaded area gives the unstable domain in the absence of collisions
$(\beta=0)$. The dashed and dotted curves show how the boundaries 
change as collisions are included. They correspond to the cases 
$\beta=.3$ (dashed) adn $\beta=1.0$ (dotted).} 
\end{figure}
\noindent we note that formation of a density grating increases 
the mean field energy. Collisions, therefore, should
join diffraction in opposing the bunching process, resulting in a higher
threshold for the instability. 
For negative scattering lengths, on the other hand, bunching reduces the mean 
field energy, hence collisions should enhance the bunching process and oppose
diffraction, thus lowering the threshold. Lastly, we note that the instability
region touches $\chi^2=0$ when $\delta=\sqrt{1+2\beta}\approx1+\beta$.
This roughly corresponds to the conservation of energy in the scattering
of a pump photon into the probe by an atom intitially at rest.

Once we have found the eigenvectors and eigenvalues of ${\bf M}$,
we can reexpress the solution (\ref{deltaoft}) in the form
\begin{equation}
\hat{\vec{\delta}}(\tau)={\bf U}e^{i{\mit \Omega}\tau}{\bf U}^{-1}
\hat{\vec{\delta}}(0),
\label{deltaeigen}
\end{equation}
where ${\bf U}$ is the matrix of eigenvectors of ${\bf M}$, such that 
$U_{ij}$ is the $i$th component of the $j$th eigenvector, and ${\mit\Omega}$ 
is the diagonal matrix of eigenvalues, such that the $i$th diagonal element of 
${\mit \Omega}$ is the $i$th eigenvalue of ${\bf M}$. 
In the unstable regime, we have the eigenvalues $\omega_1$,
$\omega_2=\Omega+i\Gamma$, and $\omega_3=\Omega-i\Gamma$, where
$\omega_1$ and $\Omega$ are real, and $\Gamma$ is real and positive.
Thus $\omega_1$ corresponds to an oscillating solution,
$\omega_2$ an exponentially decaying solution, and $\omega_3$ corresponds to
an exponentially growing solution. Eventually, this exponentially growing
solution will dominate, at which time we can neglect the other two terms,
yielding the approximate solution 
\begin{equation}
\hat{\delta}_j(\tau)=\sum_k\zeta_{jk}\hat{\delta}_k(0)e^{(\Gamma+i\Omega)\tau},
\label{deltaegr}
\end{equation}
\centerline{\psfig{figure=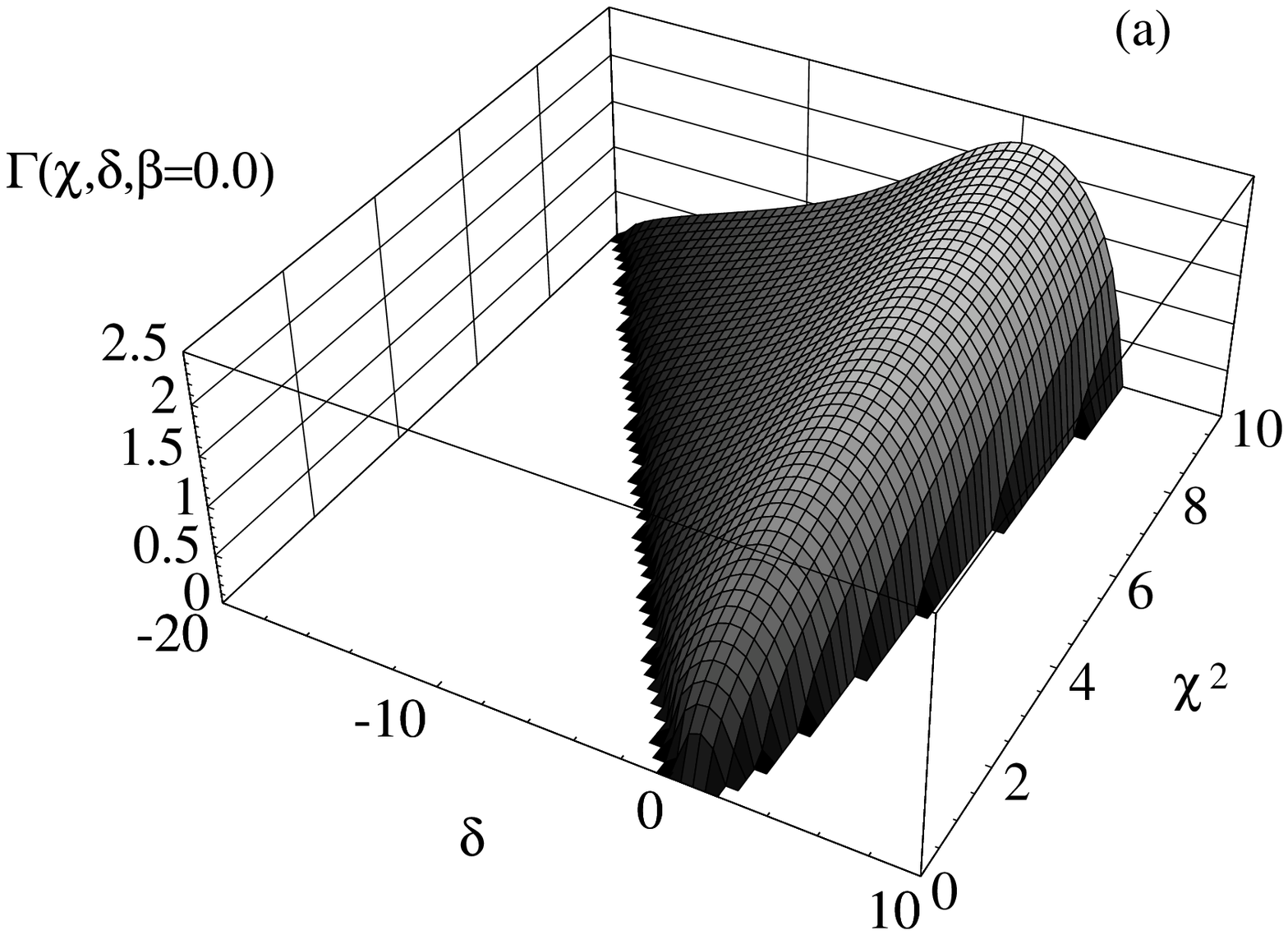,width=8.6cm,clip=}}
\centerline{\psfig{figure=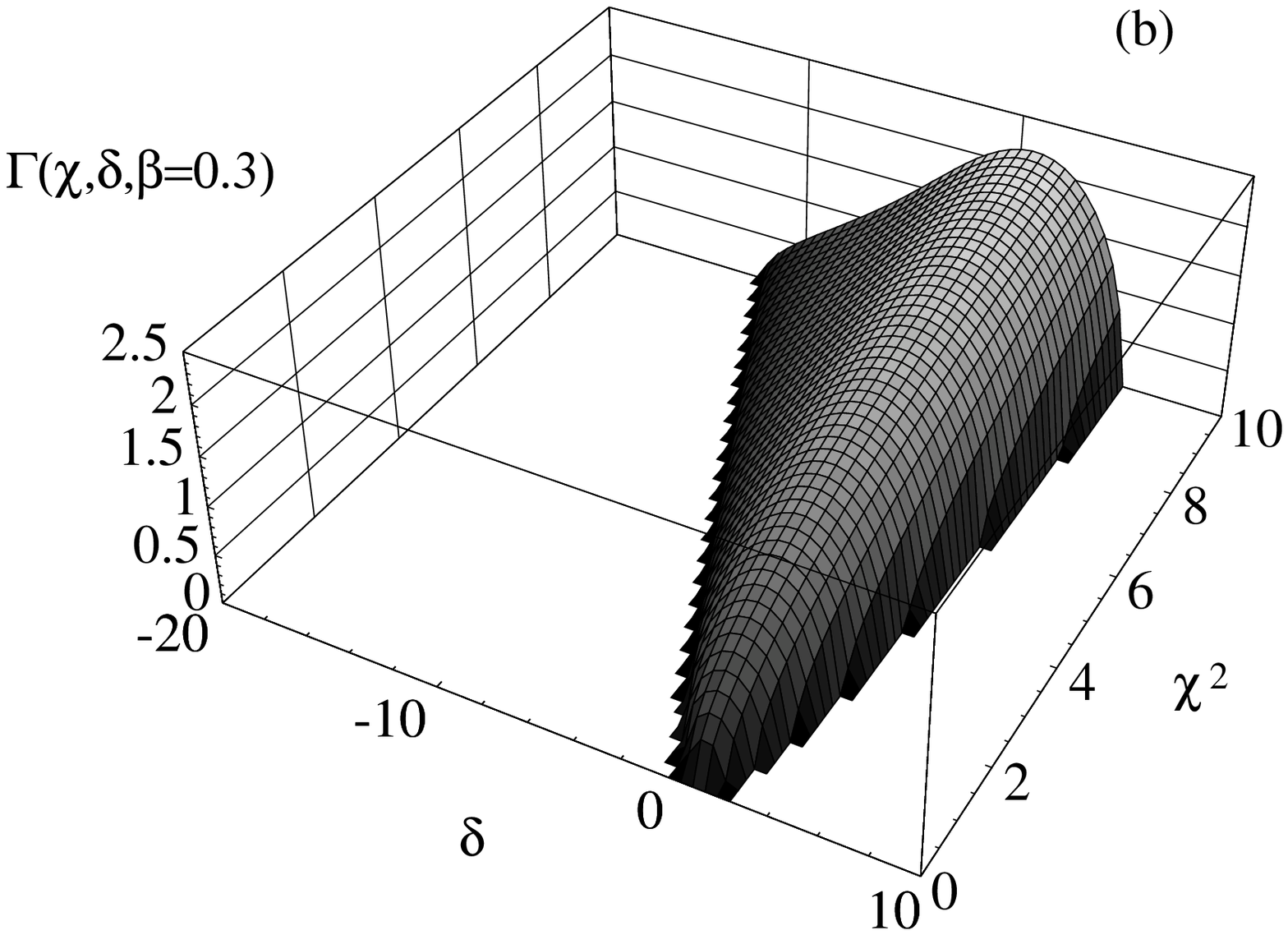,width=8.6cm,clip=}}
\centerline{\psfig{figure=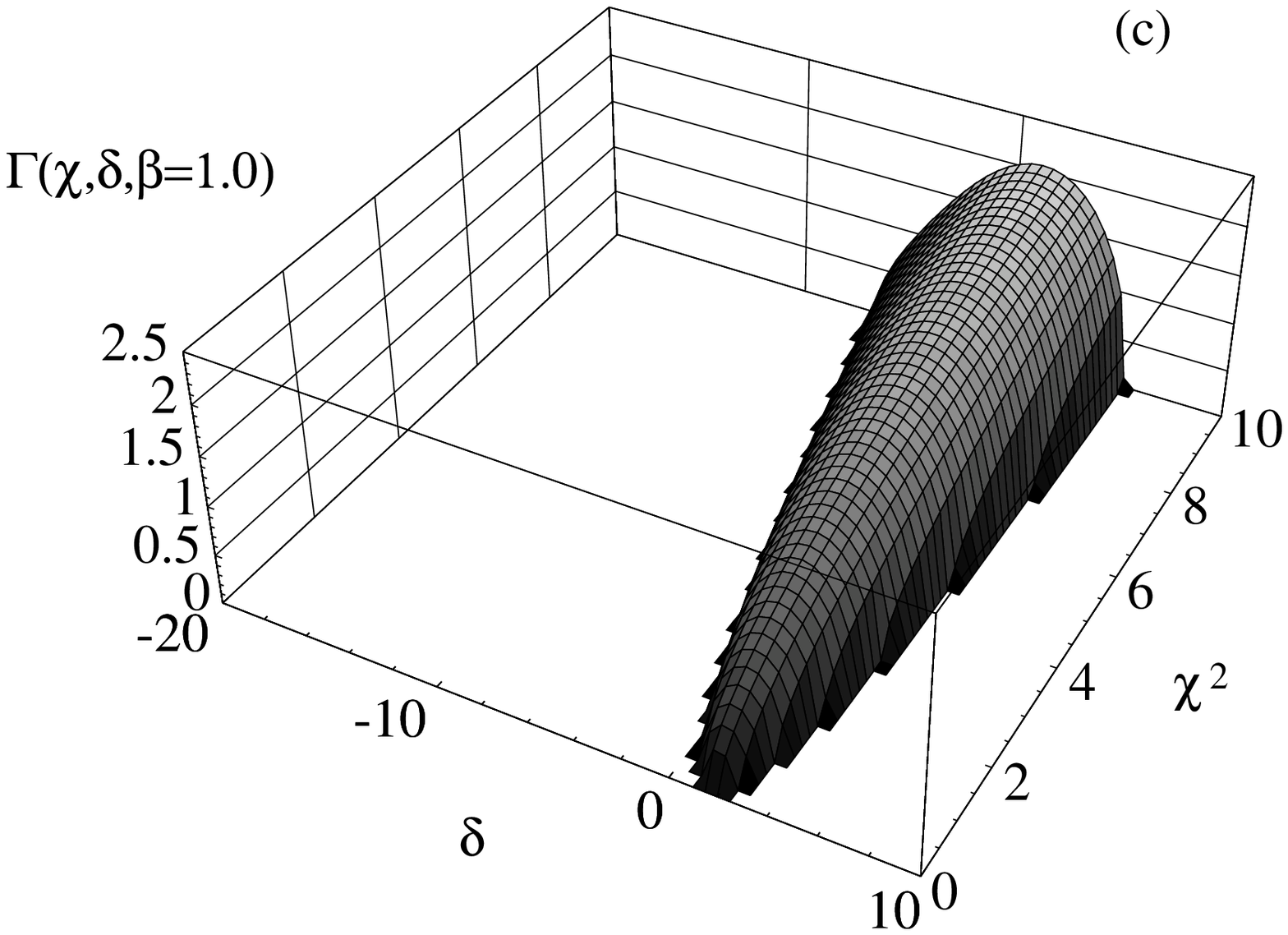,width=8.6cm,clip=}}
\begin{figure}
\caption{Exponential growth rate $\Gamma$ as a function of the scaled
pump-probe detuning $\delta$ and coupling parameter $\chi$, for various
values of the collision parameter $\beta$.
Figure 2a shows the case $\beta=0$, while Figs. 2b and 2c show the cases 
$\beta=0.3$ and $\beta=1.0$, respectively.}
\end{figure}
\noindent where $\zeta_{jk}=U_{j3}U^{-1}_{3k}$.
The range of validity for this approximation is roughly
$1<\Gamma\tau\ll\ln(\sqrt{N})$, where the lower limit is set by the requirement that
the exponentially growing terms dominate, and the 
\centerline{\psfig{figure=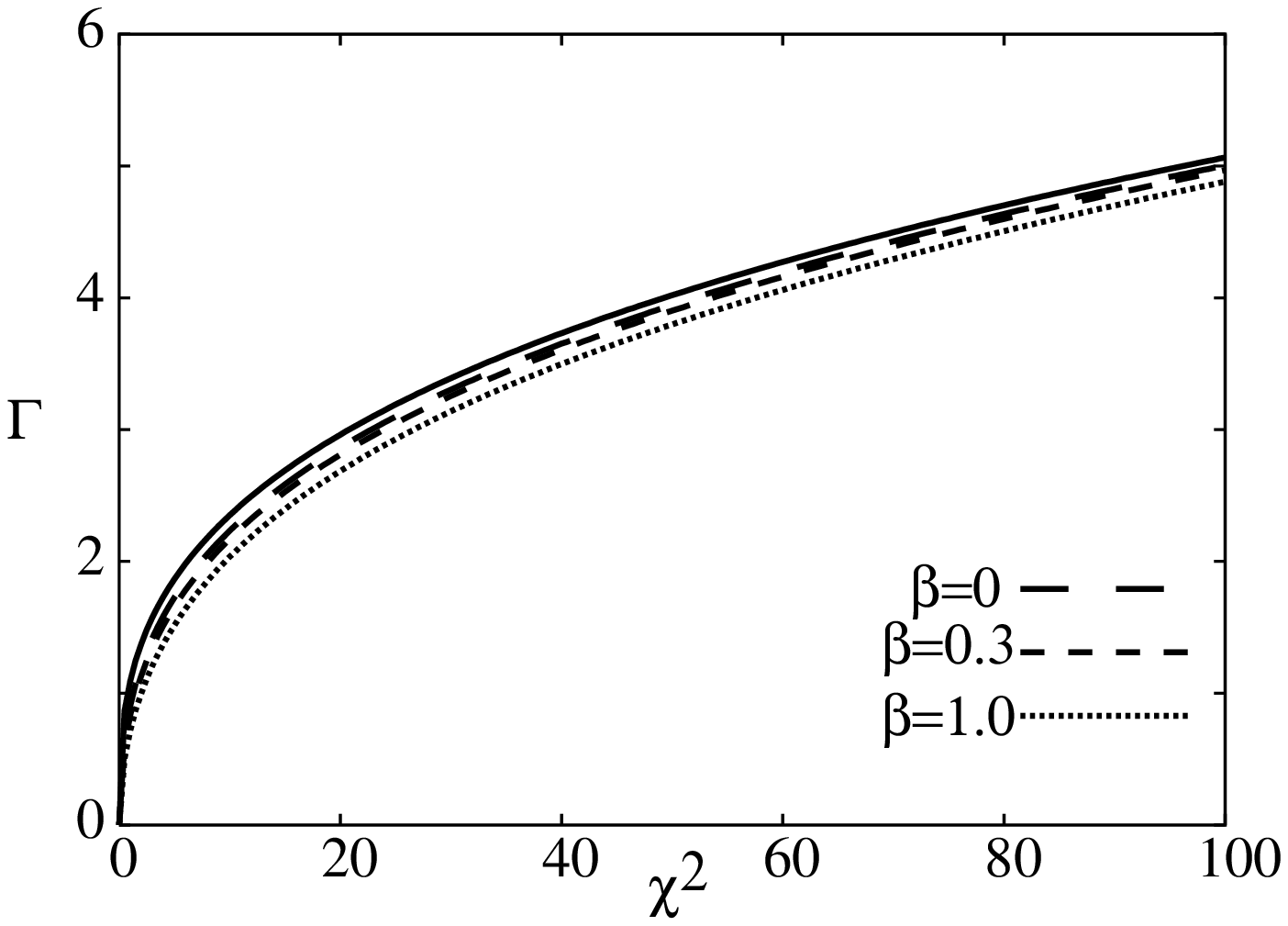,width=8.6cm,clip=}}
\begin{figure}
\caption{The dashed lines show the growth rate $\Gamma$ as a function of 
$\chi^2$ for the case $\delta=1+\beta$, with the values of $\beta$ specified in
the figure.
The solid line is the approximate expression given by Eq. (\ref{gammaest}).}
\end{figure}
\noindent upper limit comes from
the requirement that the side mode populations remain a small fraction of
the total atom number. This condition therefore formally defines the 
{\it exponential growth regime}.

The rate of exponential growth $\Gamma$ has the explicit form
\begin{equation}
\Gamma=\frac{\sqrt{3}}{2}\left|\left[r+\sqrt{q^3+r^2}\right]^{1/3}
-\left[r-\sqrt{q^3+r^2}\right]^{1/3}\right|,
\label{Gamma}
\end{equation}
where
\begin{equation}
r=-\frac{1}{3}\delta(1+2\beta)-\chi^2+\frac{\delta^3}{27},
\label{r}
\end{equation}
and
\begin{equation}
q=-\frac{1}{9}(3+6\beta+\delta^2).
\label{q}
\end{equation}
As this equation is complex and does not provide much insight, we have plotted
$\Gamma$ as a function of $\delta$ and $\chi^2$ for three different values of
$\beta$. Figure 2a shows the limit of negligible collisions $\beta=0$,
and Figs. 2b and 2c show the cases $\beta=.3$ and $\beta=1$ respectively.
From these figures, we observe that the most significant effect
of collisions is to shift the lower threshold. The values of $\Gamma$
in the vicinity of the maximum (for fixed $\chi$), on the other hand, show
less pronounced variations. 

In cases where $\chi^2 \gg |\delta^3|,|\beta|$ we have
$r\approx-\chi^2$ and $\sqrt{q^3+r^2}\approx \chi^2$.
In this case Eq. (\ref{Gamma}) reduces simply to
\begin{equation}
\Gamma\approx\sqrt{3}(\chi/2)^{2/3}.
\label{gammaest}
\end{equation}
Among other things, this shows that the gain 
scales as the number of atoms in the condensate to the $1/3$ power.
In Fig. 3, the growth rate $\Gamma$ is plotted versus $\chi^2$
with $\delta=1+\beta$, which roughly maximizes $\Gamma$
for fixed $\chi$. The three dashed curves
correspond to different values of the collision parameter $\beta$, while
the solid line gives the approximate result (\ref{gammaest}). This shows that
the approximation is a relatively accurate
estimate of the maximum gain for all values of $\chi^2$ . 

\section{Quantum statistics}

In this section we use the solution (\ref{deltaeigen}) to compute some of the
quantum statistical properties of the system. This, however, first requires a 
more detailed discussion of the physical meaning of the infinitesimal operators. 
The first, $\hat{\delta}_a=\hat{a}/\sqrt{N}$, is clearly just a rescaling of 
the photon annihilation operator. From it one can compute all properties of 
the electric field and/or the photon satistics of the probe mode.
The atomic side mode 
operators $\hat{\delta}_-=\hat{c}^\dag_-\hat{c}_0/N$ and 
$\hat{\delta}_+=\hat{c}^\dag_0\hat{c}_+/N$, however, are not simply rescalings
of atom annihilation operators. Rather, they are directly related
to the atomic density, $\hat{\rho}({\bf r})=\hat{\mit
\Psi}^\dag_g({\bf r})\hat{\mit\Psi}_g({\bf r})$.

To illustrate this point, we expand $\hat{\rho}({\bf r})$ according
to Eq. (\ref{modeexpand}) and linearize, yielding
\begin{equation}
\hat{\rho}({\bf r})=N|\varphi_0({\bf r})|^2\left[\frac{1}{2}
+e^{i{\bf K}\cdot{\bf r}}\left(\hat{\delta}_-+\hat{\delta}_+\right)+H.c.\right].
\label{density}
\end{equation}
From this expression we see that the side mode operators $\hat{\delta}_\pm$
indeed describe the appearance of a density modulation with
wavelength $2\pi/K$.  

In addition to the atomic density, one can also express the number operators 
for the side modes in terms of $\hat{\delta}_\pm$. For example, we have
after linearization
\begin{equation}
N\hat{\delta}_-\hat{\delta}^\dag_-=
\frac {\hat{c}^\dag_-\hat{c}_0\hat{c}^\dag_0\hat{c}_-}{N}  
\to \hat{c}^\dag_-\hat{c}_-\frac{(N+1)}{N}.
\label{Na}
\end{equation}
Hence with $N+1\approx N$ the number operator for the
`-' side mode can be expressed as
\begin{equation}
\hat{c}^\dag_-\hat{c}_-\approx N\hat{\delta}_-\hat{\delta}^\dag_-.
\label{N-}
\end{equation}
Similarly the number operator
for the `+' side mode is given by
\begin{equation}
\hat{c}^\dag_+\hat{c}_+\approx N\hat{\delta}^\dag_+\hat{\delta}_+.
\label{N+}
\end{equation}
From these number operators, one can therefore compute the  
number statistics of the side modes in the linear regime.

From the analytical solution (\ref{deltaeigen}) it is straightforward to 
compute the properties of the
atomic and optical fields for an arbitrary initial condition. We
focus on two conditions which appear readily accessible experimentally.
In the first one, the probe field and the atomic side modes all begin
in the vacuum state. In this case the exponential growth is triggered
by vacuum fluctuations in both the probe field and the atomic density. 
A second possible triggering mechanism involves
injecting of a weak laser field into the probe mode. Both initial situations
are investigated by assuming that the probe mode is initially in the
coherent state $\alpha$, such that $\hat{a}|\alpha\rangle=\alpha|\alpha\rangle$,
the vacuum case corresponding to
$\alpha=0$. In addition, we assume throughout that the condensate
side modes begin in the vacuum state. Hence, the initial state of the 
three-mode system can be expressed as $|\alpha,0,0\rangle$, where the first index
refers to the probe mode, and the second and third indices give the states of
the momentum side modes.

\subsection{Electric field and atomic density}
The expectation value of the operator $\hat{\delta}_a$ is sufficient to
compute the mean electric field, and likewise, the mean values of
$\hat{\delta}_-$ and $\hat{\delta}_+$ are sufficient to compute the
mean atomic
density.  We now give analytic solutions for these physical quantities
and their quantum mechanical uncertainties in the exponential growth regime, 
where all but the leading exponential terms can be safely neglected.

The electric field operator for the probe field is given by
\begin{equation}
\hat{\vec{E}}({\bf r})=\vec{\epsilon}({\bf r}){\cal E}({\bf r},\tau)
\sqrt{N}\hat{\delta}_a(\tau)+H.c.,
\label{defE}
\end{equation}
where $\vec{\epsilon}({\bf r})$ is the polarization unit vector,
and 
\begin{equation}
{\cal E}({\bf r},\tau)=-\sqrt{\frac{\hbar\omega}{2\epsilon_0}}
\frac{\Omega_0\Delta}{|\Omega_0||\Delta|}\varphi_E({\bf r})
e^{-i(\omega_0/\omega_r)\tau}
\label{defcalE}
\end{equation}
contains all constants of proportionality, the normalized spatial 
wavefunction of the probe mode $\varphi_E({\bf r})$, and the oscillation
at the pump frequency $\omega_0$. The mean electric field is obtained  
by inserting Eq. (\ref{deltaegr}), and taking the quantum mechanical 
expectation value with respect to the initial state $|\alpha,0,0\rangle$, 
which yields 
\begin{equation}
\vec{\epsilon}({\bf r})\cdot\langle\hat{\vec{E}}({\bf r})\rangle
={\cal E}({\bf r},\tau)\zeta_{aa}\alpha e^{(\Gamma+i\Omega)\tau}+c.c..
\label{meanE}
\end{equation}
This corresponds to an oscillating mean field with amplitude
\begin{equation}
E_o({\bf r})=2|{\cal E}({\bf r},\tau)||\zeta_{aa}||\alpha|e^{\Gamma\tau}.
\label{E0}
\end{equation}
From this expression, we see that there is a nonzero mean field provided
only that $\alpha\neq 0$.  We also see from Eq. (\ref{meanE})
that the mean field amplitude grows exponentially at the
rate $\Gamma$, and that its frequency is shifted by $-\Omega$ away from
the pump frequency. Its phase, on the other hand, has a somewhat complicated
dependence on the system parameters. An analytic expression for this
phase can be computed directly from Eq. (\ref{meanE}), but as we draw no
specific conclusions from it, we do not give the explicit expression here.

The variance in the electric field can also be computed
in a straightforward manner from Eq. (\ref{deltaegr}), yielding
\begin{equation}
\Delta E({\bf r})=\sqrt{2}|{\cal E}({\bf r},\tau)||\zeta_{a-}|e^{\Gamma\tau}.
\label{DeltaE}
\end{equation}
This shows that the fluctuations also grow exponentially in time, irrespective
of whether the mean field vanishes or not, and are in fact
independent of $\alpha$. Hence these fluctuations can
be attributed solely to the amplification of quantum noise, i.e. vacuum
fluctuations in the probe electric field as well as atomic density fluctuations.
While the mean field and the fluctuations both grow exponentially in time, the relative
uncertainty, on the other hand, is constant in time, given by
\begin{equation}
\frac{\Delta E({\bf r})}{E_0({\bf r})}=\frac{
f(\delta,\chi,\beta)}{\sqrt{2}|\alpha|}.
\label{DEoverE}
\end{equation}
Here, we have introduced the fluctuation function
\begin{equation}
f(\delta,\chi,\beta)=\frac{|U^{-1}_{3-}|}
{|U^{-1}_{3a}|},
\label{deff}
\end{equation} 
which can be computed directly from the eigenvectors of the matrix ${\bf M}$,
and therefore depends only on the parameters $\chi$, $\delta$, and $\beta$.
Figure 4 plots $f(\delta,\chi,\beta)$  
above the $\delta$-$\chi^2$ plane for various values of the collision 
parameter $\beta$. Figure 4a shows the limit of negligible collisions
$(\beta=0)$,
where we see that $f(\delta,\chi,0)$ is nearly flat in the vicinity
of maximum gain ($\delta\approx 1$), where it has a value somewhere
between 1 and 2. It steadily increases from this value 
as the pump-probe detuning $\delta$ moves in the negative
direction. Figures 4b and 4c show the cases $\beta=0.3$ and $\beta=1.0$
respectively. From these we see that the effect of increasing the collision
parameter is to flatten $f(\delta,\chi,\beta)$ as a function of $\chi$ and
$\delta$. 

In a similar manner, we next calculate the mean value and variance of the 
atomic density. By inserting Eq. (\ref{deltaegr}) into Eq. (\ref{density})
and taking the expectation value with respect to the initial state, we
find the expectation value of the atomic density to be
\begin{eqnarray}
\langle\hat{\rho}({\bf r})\rangle=N|\varphi_0({\bf r})|^2+\rho_0({\bf r},\tau)
\cos[{\bf K}\cdot{\bf r}+\Omega\tau+\phi]
\label{meanrho}
\end{eqnarray} 
where the amplitude $\rho_0({\bf r})$ of the density modulation is given by
\begin{equation}
\rho_0({\bf r},\tau)=\sqrt{N}|\varphi_0({\bf r})|^2
|\zeta_{-a}+\zeta_{+a}||\alpha|e^{\Gamma\tau}.
\label{rho0}
\end{equation}
Thus we see that the mean atomic density is the sum of two contributions,
the initial density of the condensate plus a density modulation which 
grows exponentially in time, provided of course that $\alpha\neq 0$. 
Together with Eq. (\ref{meanE}), this shows
that the phase symmetry of the system is broken by the phase of the injected 
field. Only in the case $\alpha=0$ does the symmetry remain unbroken. We 
note that for the atomic
side modes, this is not symmetry breaking in the commonly used sense of
non-zero mean fields. Rather, it is the mean atomic density
modulation which acquires a nonzero phase.
Note also that the mean density modulation is not stationary, as its phase is 
given 
\centerline{\psfig{figure=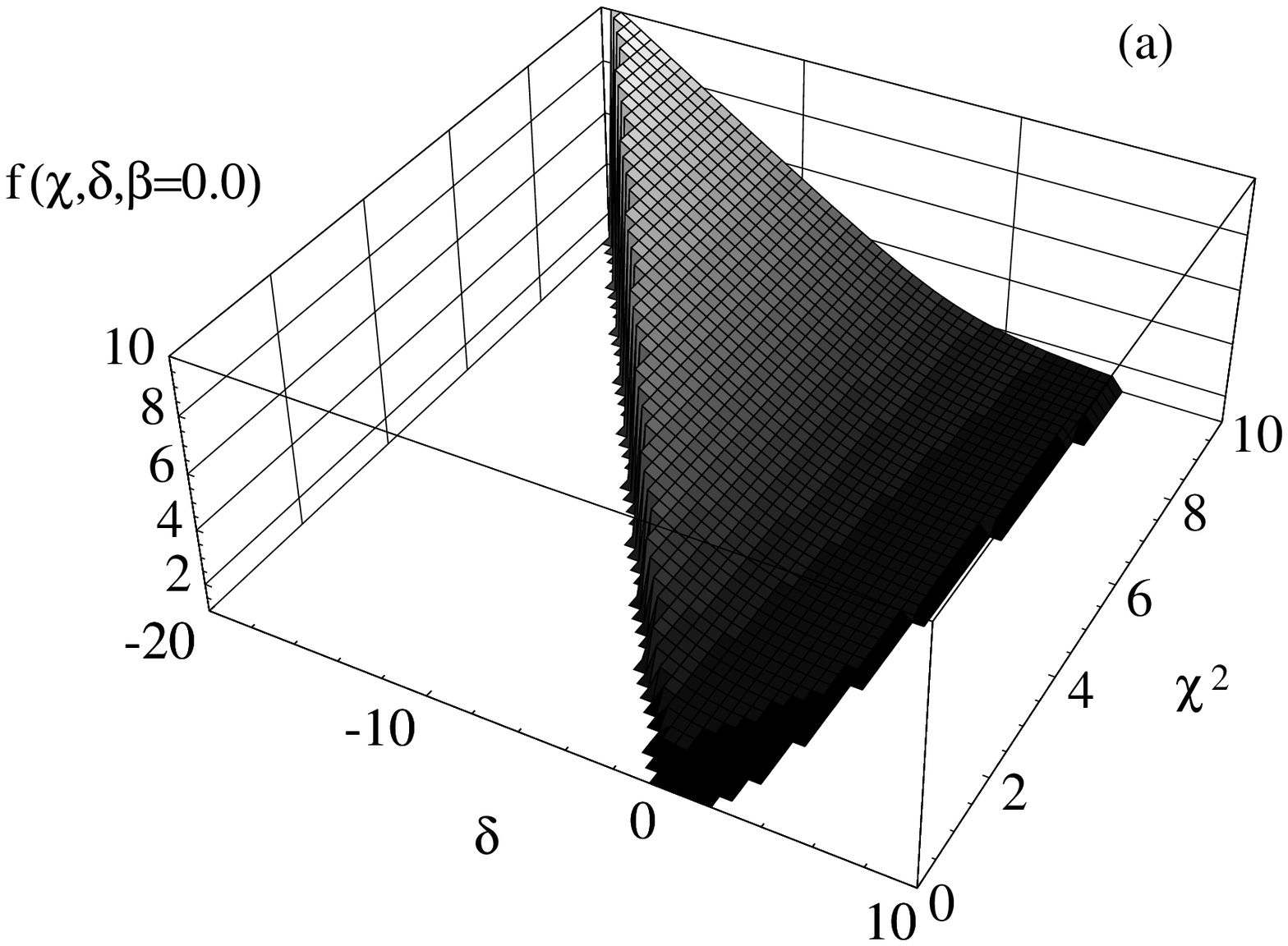,width=8.6cm,clip=}}
\centerline{\psfig{figure=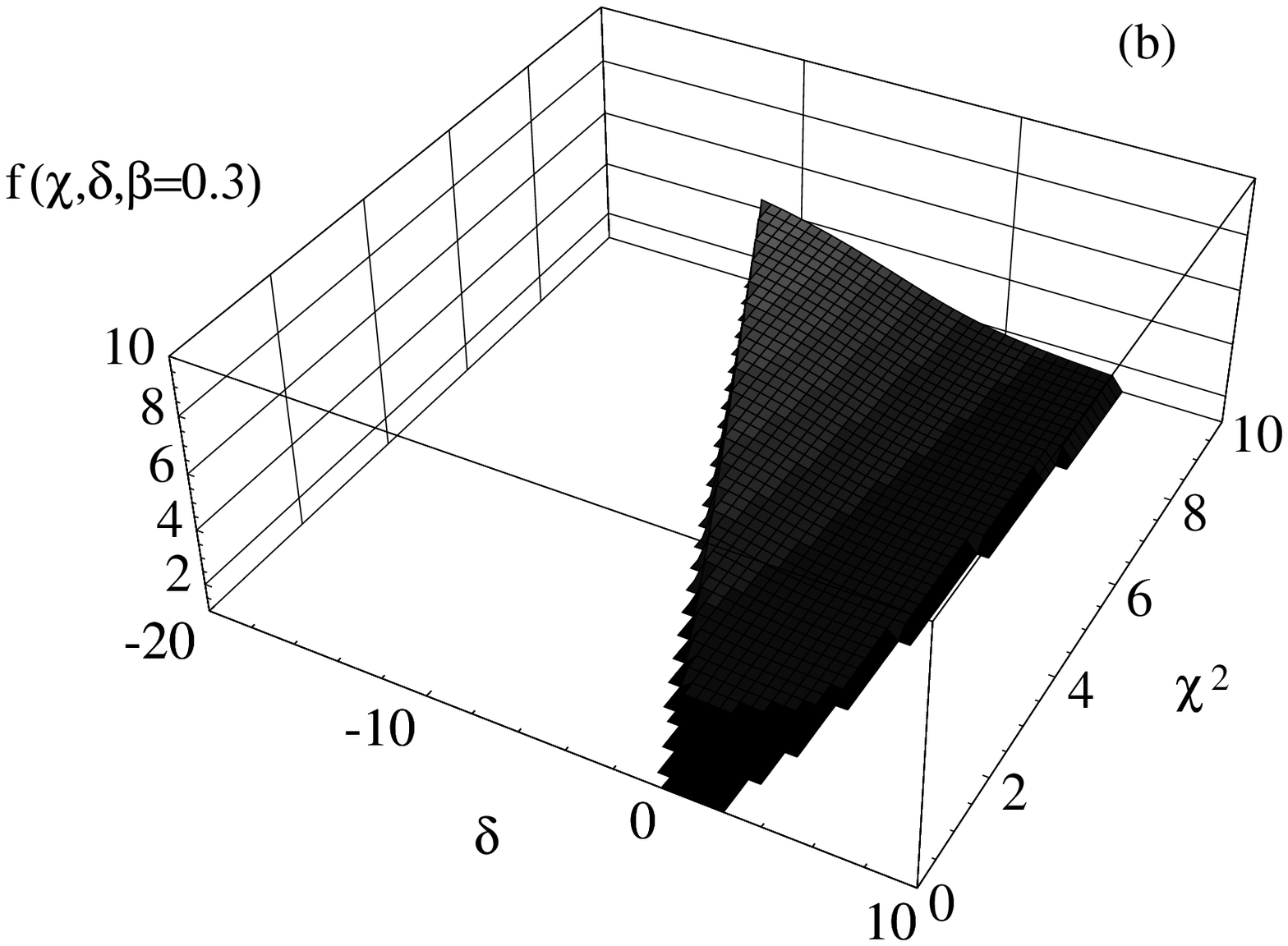,width=8.6cm,clip=}}
\centerline{\psfig{figure=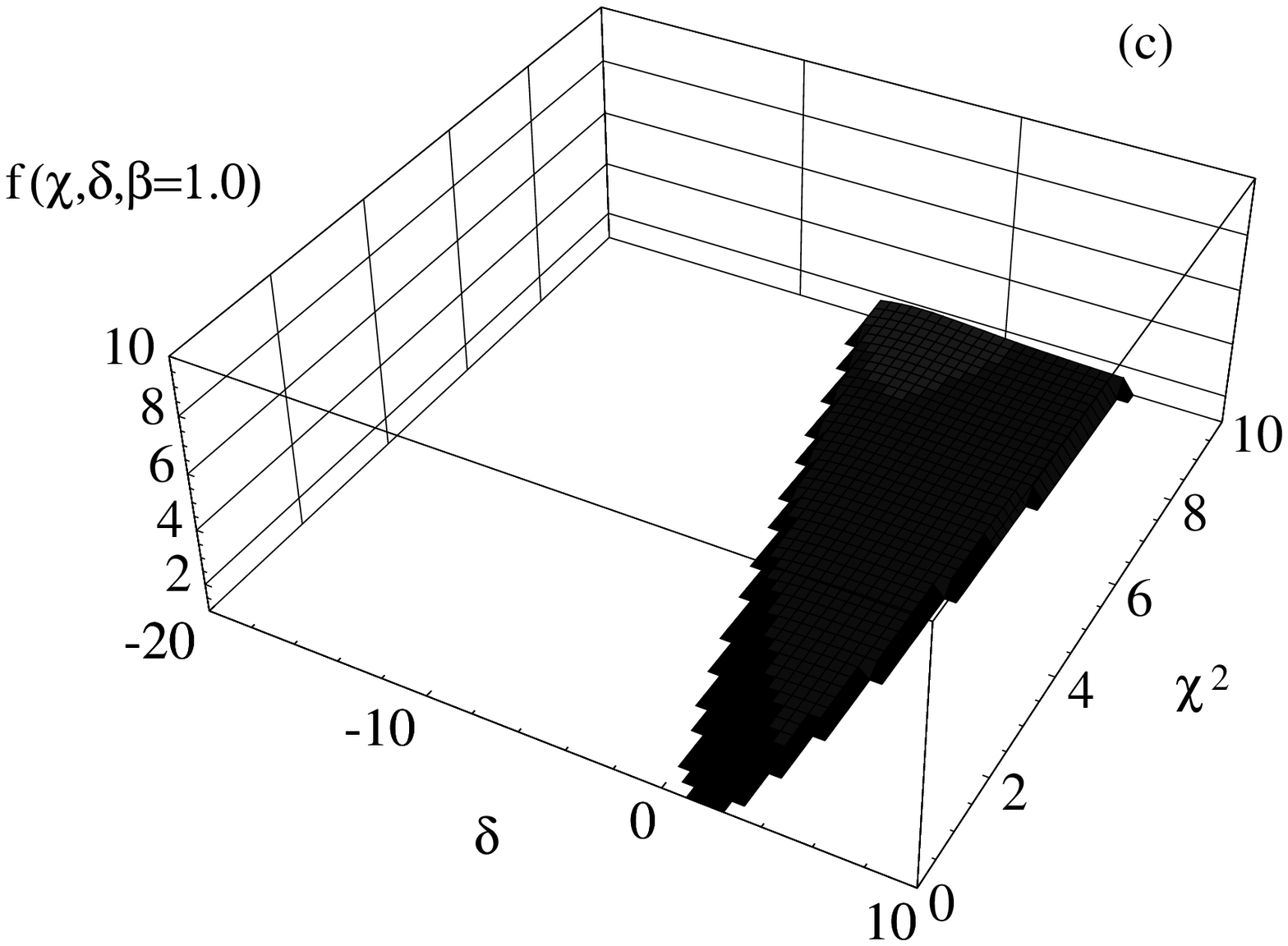,width=8.6cm,clip=}}
\begin{figure}
\caption{Comparison of the the fluctuation function $f(\delta,\chi,\beta)$
as a function of $\delta$ and $\chi^2$ for various values of the collision
parameter $\beta$. Figure 3a gives the limit of negligible collisions $(\beta=0)$,
where Figs. 3b and 3c show the cases $\beta=0.3$ and $\beta=1.0$, respectively.}
\end{figure}
\noindent by $\Omega\tau +\phi$, where both $\Omega$, and $\phi$ are depend only 
on the system parameters $\delta$, $\chi$, and $\beta$.

The variance in the atomic density can also be readily computed in the exponential
growth approximation, yielding 
\begin{equation}
\Delta\rho({\bf r})=2\sqrt{N}|\varphi_0({\bf r})|^2|\zeta_{--}+\zeta_{+-}|
e^{\Gamma\tau},
\label{Deltarho}
\end{equation}
which shows that the density fluctuations grow exponentially in time, 
even in the case $\alpha=0$. The ratio between the 
density variance and the modulation amplitude  
is constant in time, and is given by the same expression as that
for the probe, i.e.,  
\begin{equation}
\frac{\Delta\rho({\bf r})}{\rho_0({\bf r})}=
\frac{f(\delta,\chi,\beta)}{\sqrt{2}|\alpha|}.
\label{Drhooverrho}
\end{equation}
From Eqs. (\ref{DEoverE}) and (\ref{Drhooverrho}), we see that 
in the case $f(\delta,\chi,\beta)\ll |\alpha|$,
both the mean electric field and the mean atomic density modulation are 
quantum mechanically well-defined, meaning that the quantum noise is
small compared to their mean values. In this regime, both quantities
could be adequately treated as classical
(c-number) fields. Outside of this regime, however, the quantum
fluctuations play a significant role, and a classical description no 
longer suffices.

The main implication of these results is that by varying the system control 
parameters, and in particular the injected field intensity and phase, 
one can vary the mean
electric field and atomic density modulation continuously between two limits.
For $|\alpha|\ll f(\delta,\chi,\beta)$ the fields are dominated by quantum 
fluctuations,
and we can expect to find important non-classical effects. In the limit of 
a `strong' injected field, however, the fluctuations are not significant,
and the atomic and optical fields behave classically. 

\subsection{Intensities}

We now turn to the number statistics of the three field modes, concentrating on
the mean atom/photon numbers and their variances.
It is convenient to reexpress the mode number operators 
given by Eqs. (\ref{Na})-(\ref{N+}), as
\begin{equation}
\hat{N}_j=N\hat{\delta}^\dag_j\hat{\delta}_j-\delta_{j-},
\label{hatNj}
\end{equation}
where the index $j$ is again the mode label $a$, $-$, or $+$, and
the $\delta$-function accounts for the fact that
we have normally ordered the infinitesimal field operators.
It is then straightforward to derive the full time-dependent solution for 
the mean occupation numbers $N_j\equiv\langle\hat{N}_j\rangle$ as
\begin{equation}
N_j=|a_{ja}(\tau)|^2|\alpha|^2+|a_{j-}(\tau)|^2-\delta_{j-},  
\label{fullNj}
\end{equation}
where
\begin{equation}
a_{ij}(\tau)=\sum_{k=1}^3U_{ik}U^{-1}_{kj}e^{i\omega_i\tau}.
\label{defa}
\end{equation}
\centerline{\psfig{figure=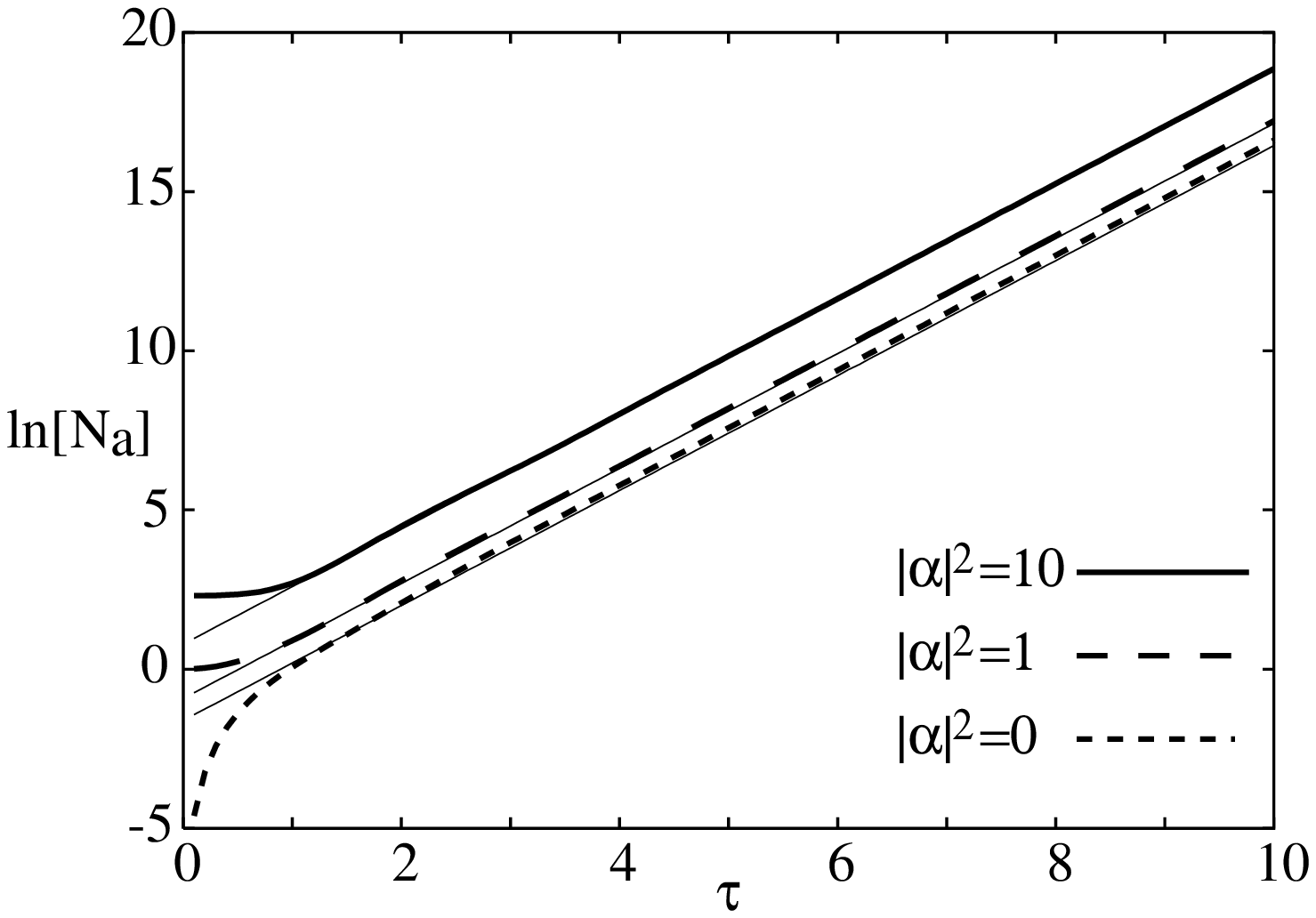,width=8.6cm,clip=1}}
\begin{figure}
\caption{Logarithmic plot of the probe intensity $N_a$ as a function of time.
This thick curves show the exact solution, given by Eq. (\ref{fullNj}),
and the corresponding thin lines give the approximate solution of Eq.
(\ref{egrNj}). The parameters chosen are $\delta=1$, $\chi^2=1$, and $\beta=0$.
Each pair of curves corresponds to a different value of of the initial probe 
intensity $|\alpha|^2$, as specified in the figure.}
\end{figure}
\noindent The first term in Eq. (\ref{fullNj}) can be interpreted as the 
stimulated contribution to the intensity,
\begin{equation}
\left[N_j\right]_{st}=|a_{ja}(\tau)|^2|\alpha|^2,
\label{defNjst}
\end{equation}
whereas the second term gives the spontaneous 
contribution,
\begin{equation}
\left[N_j\right]_{sp}=|a_{j-}(\tau)|^2-\delta_{j-},
\label{defNjsp}
\end{equation}
present even in the case $\alpha=0$.
In the exponential growth regime Eq. (\ref{fullNj}) reduces to
\begin{equation}
N_j=\left(|\zeta_{ja}|^2|\alpha|^2+|\zeta_{j-}|^2\right)e^{2\Gamma\tau},
\label{egrNj}
\end{equation}
which shows that the mode occupation grows exponentially, even in the
spontaneous case, where it was seen that the mean electric field and
mean density modulation both vanish. The validity of the exponential
approximation (\ref{egrNj}) is demonstrated in Fig. 5, where we have plotted 
the logarithm of the probe intensity as a function of time. The parameters 
chosen for the plot are 
$\delta=1$, $\chi^2=1$, and $\beta=0$. The thick lines give the full 
solution (\ref{fullNj}) for three different 
values of the initial probe intensity $|\alpha|^2=0,1,10$. The
corresponding thin lines are the approximate solutions given by Eq. 
(\ref{egrNj}), which sow good agreement for $\tau>1$. 

Turning now to quantum fluctuations in the occupation numbers, we find that
the relative uncertainties are given by
\begin{equation}
\frac{\Delta N_j}{N_j}=\sqrt{1-\frac{|a_{ja}(\tau)\alpha|^4}
{N_j^2}+\frac{1}{N_j}}.
\label{fullDNoverN}
\end{equation}
From Eq. (\ref{fullNj}) it is clear that the second term under the radicand in
the above expression is $\leq 1$, 
\centerline{\psfig{figure=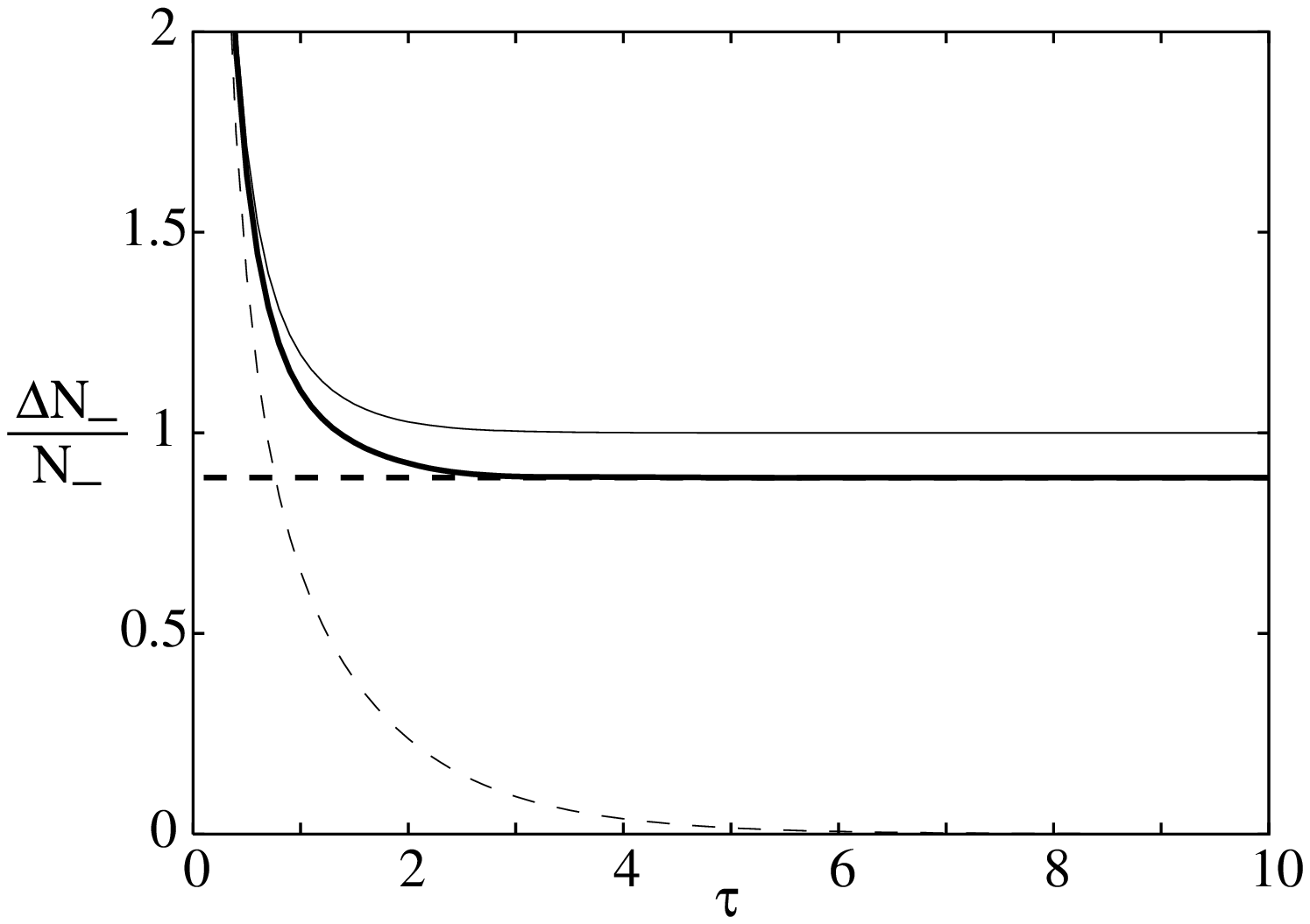,width=8.6cm,clip=1}}
\begin{figure}
\caption{The side mode number variance $\Delta N_-/N_-$ is plotted versus time 
(thick solid line). Also shown are the approximate solution given by Eq. 
(\ref{egrDNoverN}) (thick dashed line) as well as the variances for a thermal 
state (thin solid line) and a coherent state (thin dashed line) with the same 
mean value $N_-$. The parameters chosen are $\delta=1$, $\chi^2=1$, $\beta=0$, 
and $\alpha=1$.}
\end{figure}
\centerline{\psfig{figure=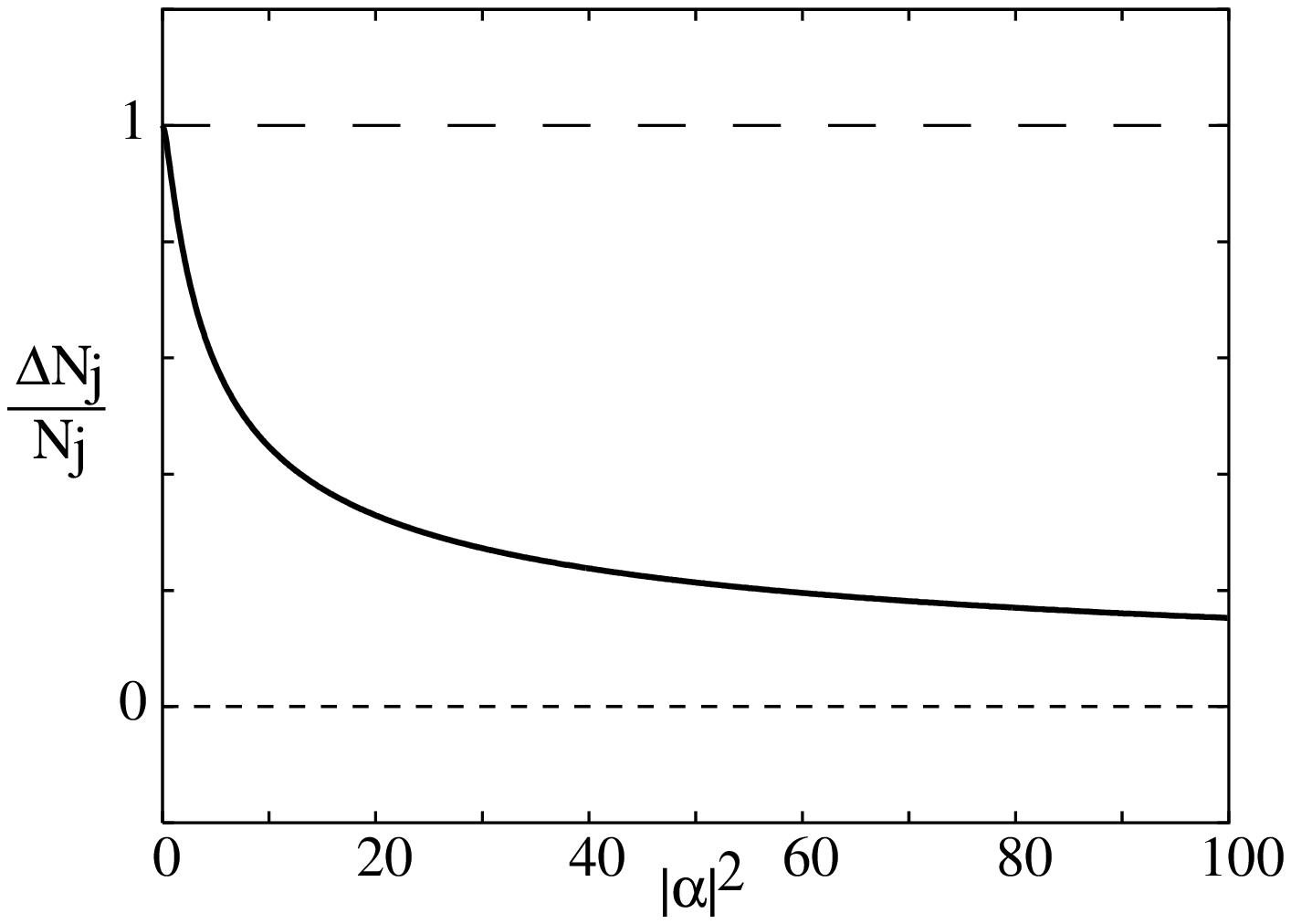,width=8.6cm,clip=}}
\begin{figure}
\caption{The long time limit of $\Delta N_j/N_j$ is plotted against the
initial probe intensity $|\alpha|^2$ (thick solid line). It various continuously
between thermal (upper dashed line) and coherent (lower dashed line) limits.
The parameters chosen are $\Delta=1$, $\chi^2=1$, and $\beta=0$.}
\end{figure}
\noindent which means that as a function of time, the relative 
uncertainty is always between $\sqrt{1/N}$, characteristic
of the fluctuations found in a coherent state, and $\sqrt{1+1/N}$, which is
the signature of thermal number fluctuations.

While for very short times the relative uncertainty may fluctuate  
between the thermal and coherent limits, once the exponentially growing terms
dominate, the relative uncertainty eventually reaches a steady-state 
value given by
\begin{equation}
\frac{\Delta N_j}{N_j}=\sqrt{1-\frac{|\alpha|^4}
{\left[|\alpha|^2+f^2(\chi,\delta,\beta)\right]^2}}.
\label{egrDNoverN}
\end{equation}
Thus we see that when $|\alpha|^2\gg f^2(\chi,\delta,\beta)$, the relative
uncertainty tends towards zero, while in the opposite case, it tends toward one.
These limits can be labeled as the
stimulated and spontaneous limits respectively. In the intermediate regime, the
fluctuations can be varied continuously between the thermal and coherent limits,
e.g. by varying the injected laser intensity, thus achieving optical control
over the quantum statistics of matter waves.

The behavior of the particle number variances is illustrated by Figs. 6 and
7. In Fig. 6 the full time dependence of $\Delta N_-/N_-$ is shown (thick solid
line). Also shown are the approximate solution given by Eq. (\ref{egrDNoverN})
(thick dashed line) as well as the variances for a thermal state (thin solid
line) and a coherent state (thin dashed line) with the same mean value $N_-$.
The parameters chosen are $\delta=1$, $\chi^2=1$, $\beta=0$, and $\alpha=1$.
Thus we see that the variance always falls between those of thermal and 
coherent fields. We also see that the long time behavior is well approximated
by Eq.
(\ref{egrDNoverN}). In Fig. 7 the steady state value for large $\tau$ is plotted
as a function of the initial probe intensity $|\alpha|^2$. Thus we see that
it is possible to vary the output continuously over the whole
range between thermal and coherent limits, simply by varying the initial probe
intensity. The parameters chosen for the figure
are $\delta=1$, $\chi^2=1$, and $\beta=0$.

\section{atom-photon entanglement}

We have previously discussed the analogy between the present system and the 
non-degenerate optical 
parametric amplifier (OPA). One of the most interesting applications of the OPA
is the generation of entangled quantum optical states. We show that
similar entanglements occur in the present system, but they are now between 
atomic and optical field modes. We first examine the two-mode intensity 
correlation functions, which give a measure of entanglement, and can be used to 
determine whether or not non-classical correlations exist between the three 
field modes. We then discuss the issue of two-mode squeezing, and show how this 
phenomenon manifests itself in the present system. 

\subsection{Two-mode intensity correlations}

The equal-time intensity correlation functions are defined in the usual manner as
\begin{equation}
g^{(2)}_{ij}=\frac{\langle\hat{N}_i\hat{N}_{j}\rangle
-\delta_{ij}\langle\hat{N}_j\rangle}
{\langle\hat{N}_i\rangle\langle\hat{N}_{j}\rangle}.
\label{defg2ij}
\end{equation}
For classical fields, the two-mode ($i \ne j$) correlations are constrained by 
the Cauchy-Schwartz inequality
\begin{equation}
g^{(2)}_{ij}\leq \left[g^{(2)}_{ii}\right]^{1/2}
\left[g^{(2)}_{jj}\right]^{1/2}.
\label{CS}
\end{equation}
\centerline{\psfig{figure=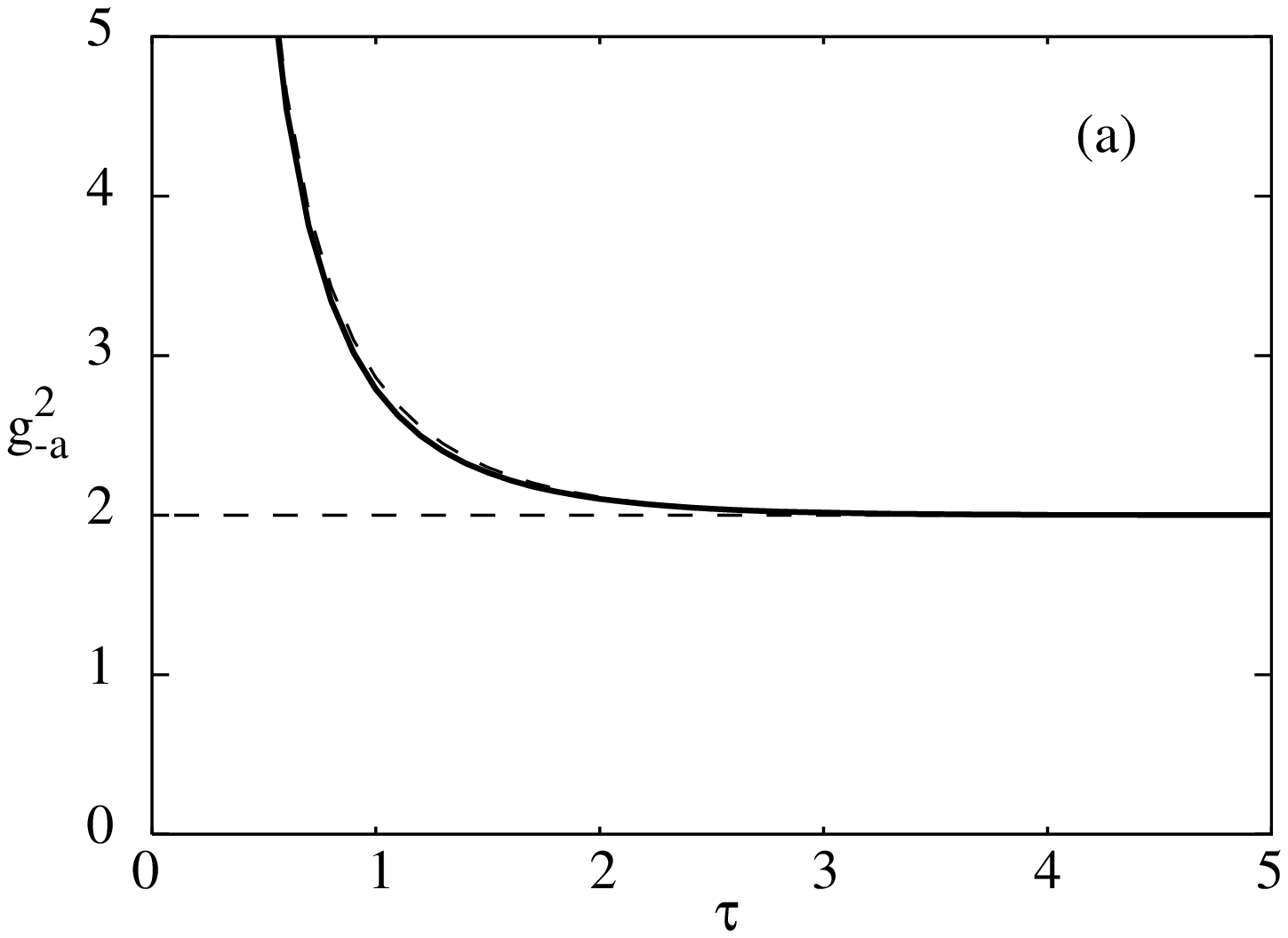,width=8.6cm,clip=1}}
\centerline{\psfig{figure=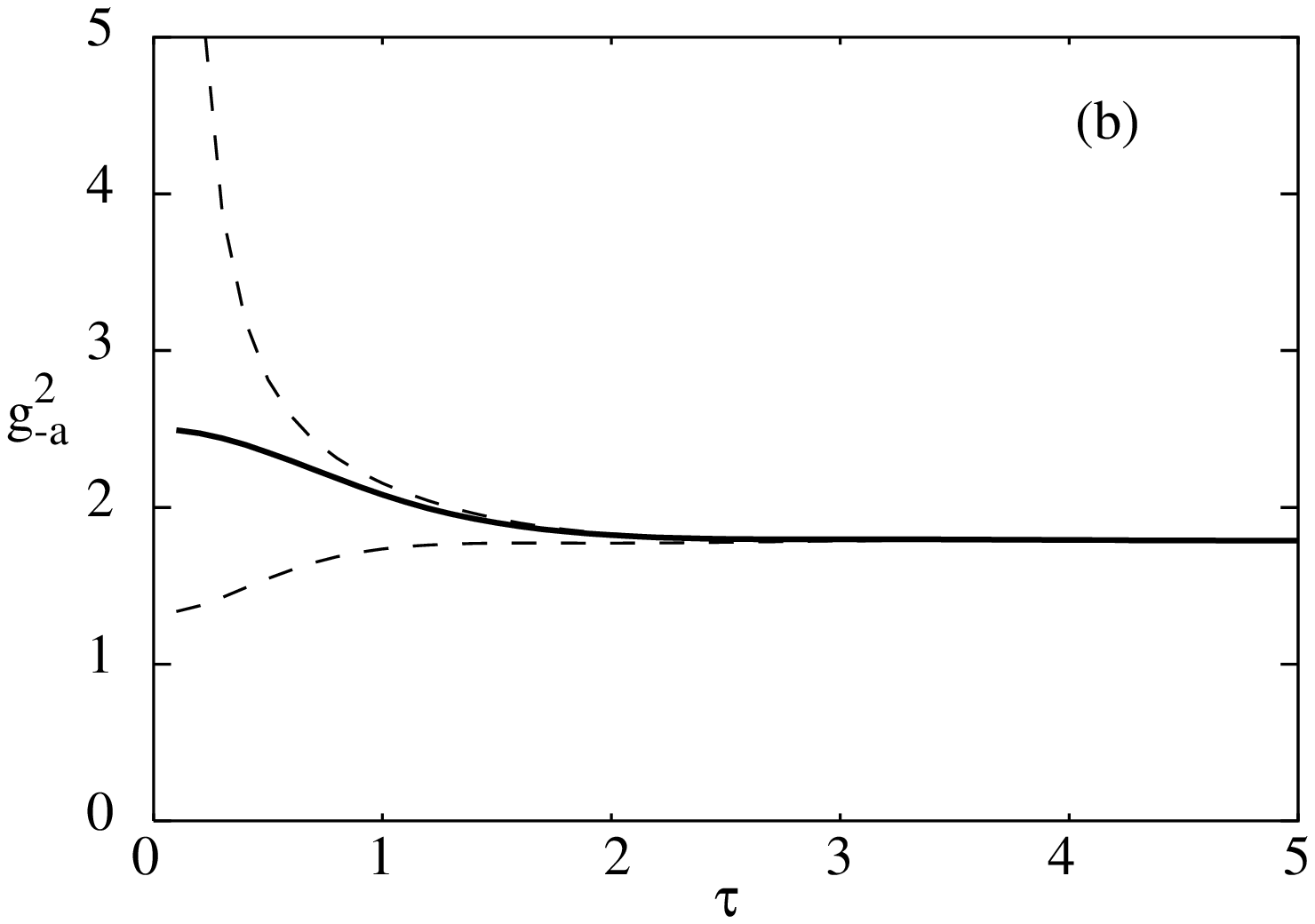,width=8.6cm,clip=1}}
\begin{figure}
\caption{The cross-correlation function $g^{(2)}_{-a}$ is plotted as a 
function of time (solid line). The upper dashed line gives the maximum allowed
by quantum mechanics, while the lower dashed line gives the upper limit for
classical correlations only. The parameters chosen are $\delta=1$, $\chi^2=1$,
and $\beta=0$. Figure 8a corresponds to triggering from noise $(\alpha=0)$,
while Fig. 8b shows the case $\alpha=1$.}
\end{figure}
\noindent Quantum mechanical fields, however, can violate this inequality
and are instead constrained by
\begin{equation}
g^{(2)}_{ij}\leq\left[g^{(2)}_{ii}+\frac{1}{\langle\hat{N}_i\rangle}\right]^{1/2}
\left[g^{(2)}_{jj}+\frac{1}{\langle\hat{N}_j\rangle}\right]^{1/2},
\label{qmineq}
\end{equation}
which reduces to the classical result in the limit of large intensities.

We focus our attention on the spontaneous case $\alpha=0$. Here 
the single-mode intensity correlation functions are those of thermal fields,
$g^{(2)}_i(\tau)=2$. In this case,
the equal-time intensity cross-correlation functions are found to be
\begin{eqnarray}
g^{(2)}_{a-}&=& g^{(2)}_{-+} =
\left[2+\frac{1}{N_a+N_+}\right]^{1/2}
 \left[2+\frac{1}{N_-}\right]^{1/2}, \nonumber \\
g^{(2)}_{a+} &=& 2.
\label{g212}
\end{eqnarray}
From Eq. (\ref{g212}) we see that both $g^{(2)}_{a-}(\tau)$ and
$g^{(2)}_{-+}(\tau)$ violate the Cauchy-Schwartz inequality, while
$g^{(2)}_{a+}(\tau)$ is consistent with classical cross-correlations.
Furthermore, the explicit evaluation of the $\zeta_{ij}$'s shows that
$I_+(\tau)\ll I_a(\tau)$, which implies that $g^{(2)}_{a-}(\tau)$ is very
close to the maximum violation of the classical inequality consistent
with quantum mechanics, whereas for $g^{(2)}_{-+}(\tau)$
the violation is not close to the allowed maximum. In the two-mode 
parametric amplifier,
the two-mode correlation function shows the maximum violation of the
Cauchy-Schwartz inequality consistent with quantum mechanics. In the three-mode
system, however, the two-mode cross-correlation functions involve a trace
over the third mode, hence it is not surprising that they
are not maximized. This is illustrated in Fig. 8a, where we have plotted
the correlation function $g^{(2)}_{-a}$ as a function of time (solid line). 
Also shown for comparison are the quantum mechanical upper limit given by Eq. 
(\ref{qmineq}) (upper dashed line), as well as the classical upper limit given
by Eq. (\ref{CS}) (lower dashed line). The parameters are $\delta=1$,
$\chi^2=1$, $\beta=0$, and $\alpha=0$. We see that when the system is
triggered by vacuum fluctuations alone, $g^{(2)}_{-a}$ is virtually
indistinguishable from the maximum allowed by quantum mechanics.

If we now allow for an injected coherent
probe field $(\alpha\neq 0)$, we first note that the intensities
are increased by approximately $|\alpha|^2$, which means that the time scale 
on which the classical and quantum upper limits (\ref{CS}) and
(\ref{qmineq}) converge is reduced by $1/|\alpha|^2$,
making an experimental confirmation of quantum correlations more difficult.
In addition, whereas
for the spontaneous case $\alpha = 0$, numerics show the cross-correlation
$g^{(2)}_{a-}$ follows almost exactly the quantum upper limit (\ref{qmineq})
for all $t>0$, for $\alpha \neq 0$, it lies somewhere in between the quantum
(\ref{qmineq}) and classical (\ref{CS}) limits. As $\alpha$ is increased,
it falls ever closer to the
classical upper limit, so that in the limit of very large $\alpha$,
the fields exhibit classical cross-correlations only. This effect is illustrated
by Fig. 8b, which is identical to 8a, except that now we have taken $\alpha=1$.
This shows that even when the probe field initially contains only one photon 
on average, it is enough to significantly reduce the quantum correlations from 
the quantum mechanical maximum.

\subsection{Two-mode squeezing}

To complete the study of atom-photon entanglement we complement the
investigation of intensity cross-correlations with a discussion of
phase-sensitive two-mode correlations. Drawing again on the similarity
between the current model and the two-mode parametric amplifier we expect
the correlations between the cavity mode and the atomic side-mode ``$-$" to
be of particular interest as a ``squeezing-like" behavior may occur. In analogy
to the parametric amplifier we thus introduce the quadrature components of
the superposed atomic and optical fields
\begin{equation}
X_{\theta}=\sqrt{\frac{N}{2}}(\hat\delta_ae^{i\theta}+\hat\delta_-e^{-i\theta}
+h.c.).
\end{equation}
The variance of $X_\theta$ is given by
\begin{eqnarray}
V(X_{\theta})&=&\langle X_{\theta}^2\rangle-\langle
X_{\theta}\rangle^2\nonumber\\
&=&|a_{a-}(\tau)e^{i\theta}+a_{--}(\tau)e^{-i\theta}|^2.
\label{phasevar}
\end{eqnarray}
Note that this variance (\ref{phasevar}) is independent of the injected signal 
strength $\alpha$, just as with the electric field, atomic density, and 
mode intensities. It follows from Eq.\ (\ref{phasevar}) that the
angle $\theta_{min}(\tau)$ that minimizes the quadrature variance is
determined by
\begin{equation}
\mbox{arg}[a_{a-}(\tau)a^\ast_{--}(\tau)]+2\theta_{min}(\tau)
=\pi
\end{equation}
so that the corresponding minimum of $V$ for fixed $\tau$ is given by
\begin{eqnarray}\label{minvar}
V_{min} (\tau)&=&\left[|a_{a-}(\tau)|-|a_{--}(\tau)|\right]^2\nonumber\\
\end{eqnarray} 
The Heisenberg uncertainty principle gives $V(X_{\theta})
V(X_{\theta+\pi/2}) \ge 1$, hence, a quadrature component
is squeezed provided $V(X_{\theta})<1$.  

In Fig. 9, we display $V_{min}(\tau)$ for various values of the system 
parameters. We find that for $\delta\approx 1$, i.e., maximum exponential 
growth rate, $V_{min}$ is a concave function of $\tau$ displaying a single 
(global) minimum
which is typically of the order of 10$^{-1}$ (cf. the full and
dashed curves). The maximum squeezing time $\tau_m$, i.e., the largest $\tau$
for which $V_{min}(\tau)=1$, is given by $\tau_m=3.5$ for $\chi=1$, $\delta=1$,
and $\beta=0$ whereas for $\chi=10$ $\tau_m$ decreases to a value of 0.20. The
reduction is due to the increase in the exponential growth rate
$\Gamma$ with larger $\chi$.
From Fig. 9, we see that squeezing does indeed occur over a broad range of
parameters, however it only persists over intermediate time scales.
For long times, $V_{min}(\tau)$ is dominated by the exponential behavior,
which eventually leads to  the violation of the squeezing condition. 
This is in contrast to the two-mode OPA, where the quadrature 
component remains squeezed for all time. 

To understand how the presence of a third mode quenches the squeezing,
we reexpress the squeezing condition  
with the help of Eq. (\ref{defNjsp}), yielding
\begin{equation}
\left(\left[N_a\right]_{sp}-\left[N_-\right]_{sp}\right)^2<2\left[N_a\right]_{sp}.
\label{squeezed}
\end{equation}
As our analysis of squeezing so far has not made explicit use of the particular
form of the matrix ${\bf M}$, it can also be applied to the standard 
two-mode optical parametric amplifier. One simply chooses a suitable ${\bf M}$ 
where the third mode is decoupled. 
\centerline{\psfig{figure=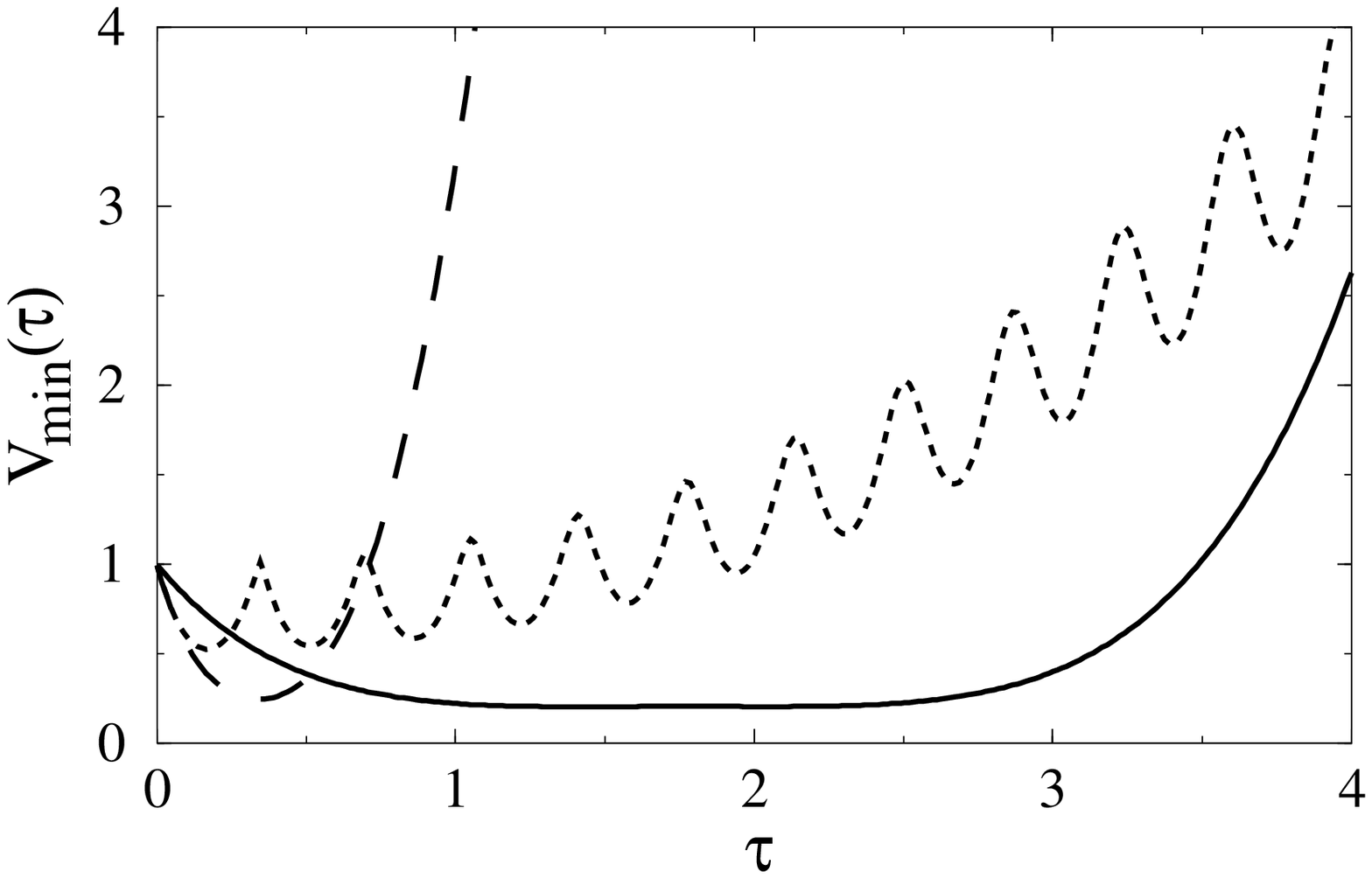,width=8.6cm,clip=1}}
\begin{figure}
\caption{Minimum variance $V_{min}$ of the quadrature components
$X_{\theta}$ as a function of $\tau$ for parameter values $\chi=1.0$,
$\delta=1.0$, $\beta=0.0$ (full curve); $\chi=3.0$, $\delta=1.0$,
$\beta=0.0$ (dashed); $\chi=3.0$, $\delta=-17.0$, $\beta=0.0$ (dotted).}
\end{figure}
\noindent In this case, the solution is well known \cite{WalMil94},
and the two modes are symmetrical, with equal populations.
This means that the l.h.s. of Eq. (\ref{squeezed}) is always zero, and the
squeezing condition is satisfied for all times. However, once the third mode is 
included, 
it introduces a small imbalance between the populations of
the two original modes, since momentum conservation
in requires that $N_a+N_+=N_-$. 
On long time scales, the l.h.s. of Eq. (\ref{squeezed}) grows
like $\exp(i4\Gamma\tau)$, whereas the r.h.s. only grows as 
$\exp(i2\Gamma\tau)$. Therefore, the squeezing condition (\ref{squeezed}) must 
eventually cease to be satisfied.

The introduction of moderate collisional interactions,
i.e., $g<1$, only leads to quantitative modifications of $V_{min}$
without altering the characteristic features. An interesting change in
behavior occurs, however, if $\delta$ is tuned closer to the borders of the
amplification range (dotted curve in Fig. 6). Under these
circumstances squeezing still occurs but $V_{min}$ displays an oscillating
behavior as a function of $\tau$. These oscillations are caused
in part by the reduction of the exponential growth rate and a simultaneous
increase of the imaginary part of the eigenvalues of the matrix 
${\bf M}$. 

The extremal angle $\theta_{min}$ varies as time evolves, eventually attaining
a constant value when the behavior of $V_{min}(\tau)$ is dominated by 
exponential growth. However, in many cases
$V(X_\theta)$ is well approximated by $V_{min}(\tau)$
if $\theta$ is chosen in the vicinity of $\theta_{min}(\tau)$.

\section{discussion and conclusion}

The fundamental time scale in the system is the inverse growth rate
$\Gamma^{-1}$. It is estimated by Eq. (\ref{gammaest}) in units of the
inverse recoil frequency $\omega_r^{-1}$ which, for sodium is given by 
$\omega_r\approx$1.7 $\mu$s. The estimated growth rate is then of the order of
$\chi^{2/3}=(\sqrt{2n_eN}g/\omega_r)^{2/3}$,
where $n_e=(I_0/8I_{sat})(\gamma/\Delta)^2$. Here $I_0$ is the pump 
intensity, $I_{sat}$ the atomic saturation intensity, and $\gamma$ the atomic
spontaneous decay rate. For sodium atoms we have $I_{sat}=6.33$mW/cm$^2$.
The parameter $n_e$ equals the fraction of excited atoms, hence, under the 
far-off 
resonance conditions we are considering, we have $n_e\ll 1$. If we chose the 
ring cavity length $L=0.1$m and effective cross section $S=10^{-9}$m$^2$,
the atom-cavity coupling constant $g$ is of the order 10$^6$ s$^{-1}$ so
that $g/\omega_r\approx 1$. As one has a great latitude in choosing the
values of $n_e$ and the total number $N$ of atoms in the BEC it should
be possible to vary the exponential growth rate over a wide range.

Our choice of $n$ and $n_e$ is constrained, however, by the requirement
that the spontaneous heating rate ${\cal L}$ be much smaller than 
the exponential growth rate $\Gamma$, so that spontaneous emission
can in fact be neglected. The heating rate (in units of $\omega_r$) is given 
by ${\cal L}=n_e\gamma/\omega_r$. As $\gamma/\omega_r\approx 10$
the condition $\Gamma\gg {\cal L}$ translates into $N\gg 10^3$ $n_e^2$.
As $n_e\ll 1$, this condition is practically always fulfilled.
Also related to spontaneous emission is the two-body
dipole-dipole interaction, which acts in addition to ground-state collisions.
For very cold atoms whose de Broglie wave length is large in
comparison to the pump laser wavelength the dipole-dipole interaction
can be approximated as a contact interaction \cite{HolAud97}. Comparing
the strength of this interaction to that ground-state collisions one
finds that the former is negligible under the condition
$n_e\ll 8\hbar k_0^3\sigma/m\gamma$,
which translates to $n_e\ll 10^{-3}$ for sodium atoms. If this condition is not
met the dipole-dipole interaction can still be accounted for to a good
degree of approximation by modifying the scattering length according to
$\sigma\to\sigma-I_0\gamma^3/32I_{sat}\Delta^2\omega_r$.

Another important parameter is the collision parameter
$\beta$. Estimating the quantity $NF_0$ to be of the order of the
atomic density we find the collision parameter $\beta$ to lie in the
range 0.1$-$1 for 'typical' densities around 10$^{15}$ cm$^{-3}$. If $\beta$
could be measured by, e.g. observing the boundaries of the instability regime,
than this could be used as a novel means to determine the atomic s-wave 
scattering length. 

In a realistic optical cavity, the lifetime of a photon is of the order of 
1-10ns, corresponding to a decay rate of 10$^3$ to 10$^4$ in units of the 
recoil frequency. This tells us how large the growth rate $\Gamma$ would have 
to be to result in a buildup of photons in the cavity. Since $\Gamma$ goes like
$N^{1/3}$ we would likely need a very large condensate, with 10$^{12}$ or more 
atoms, to achieve this. In the future, larger condensates and better cavities,
will be available, at which point the theory could presumably be 
tested. Future research will study the quantum statistics of the system
under the influence of probe field damping, which will relate the model
more closely to current experiments. For example, recent experiments by W.
Ketterle's group involving a condensate driven by a far-off resonant pump laser
have demonstrated the apperance of distinct momentum side modes as a 
consequence of spontaneous emission. While these experiments involve at most 
one spontaneous photon at a time in the condensate, they clearly demonstrate
many aspects of the theory we have described, as wll as
the importance of continuing to develop a nonlinear/quantum 
optics approach to the theory of optically driven Bose-Einstein condensates.
 
\acknowledgements
This work is supported in part
by the U.S. Office of Naval Research Contract No. 14-91-J1205, by the 
National Science Foundation Grant PHY98-01099, by the U.S. Army Research 
Office and by the Joint Services Optics Program.

\section*{APPENDIX A: Non-orthogonality}

By properly taking into account the non-orthogonality of the atomic field modes,
it can be shown that the only surviving effect in the linearized theory 
is the  modification of the atomic polarization term in the equation of motion
for the probe field (\ref{dadt}) to include a second scattering mechanism
in which a condensate scatters a photon without changing its center of mass
state. As a consequence of momentum conservation, this process is supressed
by a factor $\langle\varphi_0|\varphi_-\rangle$ relative to
the process which transfers the atom to the side mode state. Bose enhancement,
on the other hand, is stronger for this transition by a factor $\sqrt{N}$,
because we now have $N$ identical bosons in both the initial and final states.
Thus it is the product $\sqrt{N}\langle\varphi_0|\varphi_-\rangle$ which
must be negligible if we are to make the orthogonality approximation. 

More precisely, we find that the quantum statistics must be modified 
by taking $\alpha\to\alpha-s$, where
\begin{equation}
s=\frac{\chi\varphi_0|\varphi_-\rangle\sqrt{N}}{(\Omega-i\Gamma)},
\label{defs}
\end{equation}
in order to account for this additional scattering mechanism.
Thus when $\alpha$ is comparable to $f(\delta,\chi,\beta)$, the condition
$|s|\ll|\alpha|$ allows us to neglect $s$. 
For the case $\alpha\ll f(\delta,\chi,\beta)$, one the other hand, requires
$|s|\ll f(\delta,\chi,\beta)$.Since $f(\delta,\chi.\beta)$ is typically of order
1, then the condition $|s|\ll 1$ is sufficient to satisfy both conditions.  

For a dilute condensate, the ground state wave function $\varphi_0(\bf r)$ is 
given to a approximately by the single-particle ground state. For a harmonic 
trap with ${\bf K}$ taken along the $z$-axis, one has thus
\begin{equation}
\langle\varphi_0|\varphi_-\rangle=e^{-\frac{1}{4}(KW_c)^2},
\label{dilute}
\end{equation}
we $K$ approximately twice the optical wave number, and $W_c$ is the 
condensate width along ${\bf K}$. For a dilute condensate which is
an order of magnitude or larger than the optical wavelength, this
integral is clearly vanishingly small, and $s\approx 0$ for all reasonable
values $N$ and the system parameters $\delta$, $\chi$, and $\beta$.

In the case of a dense condensate the density distribution can be described 
with the help of the Thomas-Fermi approximation, i.e.,
\begin{equation}
|\phi_0({\bf r})|^2=[\mu-V_g({\bf r})]m/(4\pi\hbar^2\sigma N),
\label{tfe}
\end{equation}
where the chemical potential 
$\mu$ is determined
from the normalization requirement. Starting from 
Eq.\ (\ref{tfe}) and choosing the recoil wave vector ${\bf K}$ to lie parallel 
the $z$-axis, the overlap integral is found to be
\begin{equation}
\langle\varphi_0|\varphi_-\rangle=15 j_2(K W_c)/(K W_c),
\label{dense}
\end{equation}
where $j_2$ is the modified Bessel functio, and the $W$ is the condensate 
width along ${\bf K}$. For condensates which are large compared to the optical
wavelength, Eq. (\ref{dense}) behaves like 
$-15\sin(KW_c)/(KW_c)^3$. So assuming that $KW_c=100$, which corresponds
to a 10$\mu$m condensate, we would need $\chi\sqrt{N}/(\Omega-i\Gamma)$
to be of the order 10$^6$ in order achieve an appreciable value of $s$,
and this is much larger than is currently feasible.
However, if, for example, ${\bf K}$ is oriented parallel to 
a trap axis with tight confinement $KW_c$ need not be too large and 
diffraction effects may be appreciable.

A more efficient way to increase $s$ consists in confining the condensate in
three-dimensional rectangular trap potential. In this case the Thomas-Fermi 
approximation gives $|\varphi_0({\bf r})|^2=1/V$, $V$ being the volume of the 
trap. With ${\bf K}$ chosen along the $z$-axis of the trap, and $W_c$ being
the width of the box in this direction, one obtains for the overlap 
integral
\begin{equation}
\langle\varphi_0|\varphi_-\rangle=e^{-iKW_c/2}\frac{\sin(KW_c/2)}{KW_c/2}.
\end{equation}
For large $KW_c$ $I$ now decays only as $(KW_c)^{-1}$. 

\section*{APPENDIX B: Effective Hamiltonian}
If we define 
$\hat{C}_-=\sqrt{N}\hat{\delta}^\dag_-$, and $\hat{C}_+=\sqrt{N}\delta_+$,
then these operators obey approximately bosonic commutation relations, and the 
system of equations (\ref{ddeltadt}) can be derived from the effective 
Hamiltonian  
\begin{eqnarray}
{\cal H}_{eff}&=&(1+\beta)\hat{C}^\dag_-\hat{C}_-
+(1+\beta)\hat{C}^\dag_+\hat{C}_+
-\hat{a}^\dag\hat{a}\nonumber\\
&+&\beta(\hat{C}^\dag_-\hat{C}^\dag_++\hat{C}_-\hat{C}_+)\nonumber\\
&+&\chi(\hat{a}^\dag\hat{C}^\dag_-+\hat{a}\hat{C}_-+\hat{a}^\dag\hat{C}_+
+\hat{a}\hat{C}^\dag_+).
\label{Heff}
\end{eqnarray}

\end{document}